\documentclass{aastex63}
\usepackage{float}
\makeatletter
\let\newfloat\newfloat@ltx
\makeatother
\usepackage{bm}
\usepackage{csquotes}
\usepackage{physics}
\usepackage{verbatim}
\usepackage{amsmath}
\usepackage{mathrsfs}
\usepackage{algorithm}
\usepackage{algorithmic}
\usepackage{xcolor}
\accepted{\today}
\submitjournal{ApJ}

\shorttitle{Mode-Selective Photonic Lantern Nuller}
\shortauthors{Xin et al.}
\graphicspath{{./}{figures/}}

\begin{document}

\title{Efficient detection and characterization of exoplanets within the diffraction limit: nulling with a mode-selective photonic lantern}

\correspondingauthor{Yinzi Xin}
\email{yxin@caltech.edu}

\author[0000-0002-6171-9081]{Yinzi Xin}
\affiliation{Department of Astronomy, California Institute of Technology, Pasadena, CA, 91125, USA}

\author[0000-0001-5213-6207]{Nemanja Jovanovic}
\affiliation{Department of Astronomy, California Institute of Technology, Pasadena, CA, 91125, USA}

\author[0000-0003-4769-1665]{Garreth Ruane}
\affiliation{Jet Propulsion Laboratory, California Institute of Technology, 4800 Oak Grove Drive, Pasadena, CA, 91109, USA}

\author{Dimitri Mawet}
\affiliation{Department of Astronomy, California Institute of Technology, Pasadena, CA, 91125, USA}

\author[0000-0002-0176-8973]{Michael P. Fitzgerald}
\affiliation{Department of Physics \& Astronomy, 430 Portola Plaza, University of California, Los Angeles, CA 90095, USA}

\author{Daniel Echeverri}
\affiliation{Department of Astronomy, California Institute of Technology, Pasadena, CA, 91125, USA}

\author[0000-0001-8542-3317]{Jonathan Lin}
\affiliation{Department of Physics \& Astronomy, 430 Portola Plaza, University of California, Los Angeles, CA 90095, USA}

\author[0000-0002-5606-3874]{Sergio Leon-Saval}
\affiliation{Sydney Astrophotonic Instrumentation Laboratory, School of Physics, The University of Sydney, Sydney, NSW 2006, Australia}

\author[0000-0002-1955-2230]{Pradip Gatkine}
\affiliation{Department of Astronomy, California Institute of Technology, Pasadena, CA, 91125, USA}

\author[0000-0003-1392-0845]{Yoo Jung Kim}
\affiliation{Department of Physics \& Astronomy, 430 Portola Plaza, University of California, Los Angeles, CA 90095, USA}

\author[0000-0002-8352-7515]{Barnaby Norris}
\affiliation{Sydney Institute for Astronomy, School of Physics, Physics Road, The University of Sydney, NSW 2006, Australia}

\author{Steph Sallum}
\affiliation{Department of Physics \& Astronomy, University of California, Irvine, 4129 Frederick Reines Hall, Irvine, CA 92697 USA}



\begin{abstract}
Coronagraphs allow for faint off-axis exoplanets to be observed, but are limited to angular separations greater than a few beam widths. Accessing closer-in separations would greatly increase the expected number of detectable planets, which scales inversely with the inner working angle. The Vortex Fiber Nuller (VFN) is an instrument concept designed to characterize exoplanets within a single beam-width. It requires few optical elements and is compatible with many coronagraph designs as a complementary characterization tool. However, the peak throughput for planet light is limited to about 20\%, and the measurement places poor constraints on the planet location and flux ratio. We propose to augment the VFN design by replacing its single-mode fiber with a six-port mode-selective photonic lantern, retaining the original functionality while providing several additional ports that reject starlight but couple planet light. We show that the photonic lantern can also be used as a nuller without a vortex. We present monochromatic simulations characterizing the response of the Photonic Lantern Nuller (PLN) to astrophysical signals and wavefront errors, and show that combining exoplanet flux from the nulled ports significantly increases the overall throughput of the instrument. We show using synthetically generated data that the PLN detects exoplanets more effectively than the VFN. Furthermore, with the PLN, the exoplanet can be partially localized, and its flux ratio constrained. The PLN has the potential to be a powerful characterization tool complementary to traditional coronagraphs in future high-contrast instruments.

\end{abstract}



\section{Introduction} \label{sec:intro}

Exoplanet exploration was identified by the Decadal Survey on Astronomy and Astrophysics 2020 as one of the top scientific priorities; in particular, the identification and characterization of Earth-like planets will play a key role in the search for biochemical signatures of life in the universe \citep{NRC_2020Decadal}. The survey identified high-contrast imaging and spectroscopy as cornerstones for the future of exoplanet science, and prioritized a coronagraphic instrument on a flagship space mission, along with development of the US Extremely Large Telescope (ELT) program. It recommended that a large ($\sim 6$ m diameter) Infrared/Optical/Ultraviolet (IR/O/UV) space telescope with capabilities for coronagraphic spectroscopy be the first mission to enter the Great Observatories Mission and Technology Maturation Program \citep{NRC_2020Decadal}. Meanwhile, telescopes on the ground can characterize young giant planets, and have the potential to reach reflected light planets around M stars for the first time \citep{ruane_2019_spie}. 

While conventional coronagraphs dramatically reduce the photon noise from the star, they are practically limited to angular separations greater than a few $\lambda/D$ (the size of a resolution element, where $\lambda$ is the wavelength and $D$ the telescope diameter). The ability to access closer-in exoplanets would greatly increase the expected yield of detectable planets, since yield scales approximately inversely with the inner working angle (IWA), with yield $\propto$ IWA $^{-0.98}$ \citep{Stark_2015}. Additionally, planets observable with coronagraphy in the visible and near-infrared regime may fall within the inaccessible inner working angle at longer wavelengths, where features of key biosignatures such as carbon monoxide and methane exist. Gaining access to closer separations at those longer wavelengths will thus enable better characterization of planets detected. 

Meanwhile, techniques such as nonredundant masking interferometry \citep{nrm_tuthill} or cross-aperture nulling interferometry \citep{bracewell_1978,pfn} can access very small angular separations. However, these approaches result in lower efficiency than coronagraphy since only a small portion of the aperture is used. The Vortex Fiber Nuller (VFN) is an instrument concept that straddles the space between the two approaches, with a smaller IWA than coronagraphs but more efficient at routing the planet light to a diffraction-limited spectrograph than single-baseline cross-aperture interferometry \citep{Ruane2018_VFN}. This technique is capable of characterizing exoplanets within 1 $\lambda/D$, requires few optical elements, and is compatible with many coronagraph designs as a complementary characterization tool.

\begin{figure*}[t]
\begin{center}
    \includegraphics[scale=0.25]{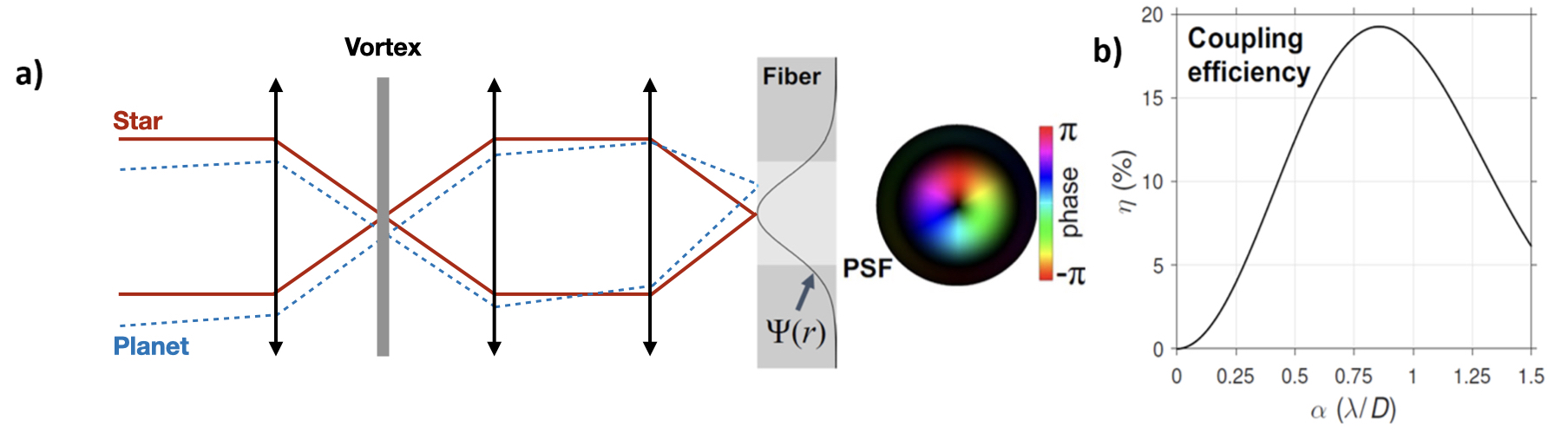}
    \caption{(a)~ Schematic of a focal-plane VFN with a single-mode fiber. The beam is focused onto a vortex mask, which imparts a different phase pattern on the star and planet point-spread-functions. The beam is then collimated and refocused onto a single-mode fiber. The on-axis star light rejected while the planet light gets partially coupled. (b)~Coupling efficiency, $\eta$, or throughput, of a planet as a function of its angular separation from the star.}
    \label{fig:fulldiagram}
\end{center}
\end{figure*}

\section{Concept}

\subsection{Vortex Fiber Nulling}
The Vortex Fiber Nuller is an instrument concept that enables spectroscopy of exoplanets within 1  $\lambda/D$, using a vortex mask to generate a vortex phase pattern on the incoming beam \citep{Ruane2018_VFN}. Figure~\ref{fig:fulldiagram}(a) shows that when the beam is on-axis (such as light from a star), the resulting pattern is orthogonal to the fundamental mode of a single-mode fiber (SMF) and does not couple to it. This result can be demonstrated by calculating the coupling efficiency of a field $f(r,\theta)$ with the SMF mode $\psi_{01}(r)$:

\begin{equation} \label{eq:overlap_integral}
    \int \psi_{01}(r) f(r,\theta) dA.
\end{equation}

For the field created by a vortex, the integral is separable, and the polar term is given by:

\begin{equation}
    \int_0^{2\pi} \exp(il\theta) d\theta,
\end{equation}

where $l$ is an integer that denotes the vortex charge. This integral evaluates to 0 for $l \neq 0$, reflecting that the vortex field is orthogonal to the SMF mode.

However, as shown in Fig.~\ref{fig:fulldiagram}(b), off-axis planet light from $\sim 0.5 \lambda/D$ to $\sim 1.3 \lambda/D$ can couple in, with a peak throughput of $19 \%$ at 0.9 $\lambda/D$. The coupled planet light can thus be directed to a spectrograph for immediate characterization, while the starlight is rejected. A focal-plane VFN is explored in this work, but \citet{ruane_2019_spie} showed that the vortex can also be placed in the pupil plane, resulting in a pupil-plane VFN that operates on the same principle of rejecting on-axis starlight with an imprinted vortex.

The range of angular separations probed by the VFN is smaller than the inner working angle of all classical coronagraphs, and is a region known to harbor potentially habitable exoplanets detected via radial velocity (RV) and transit methods. Additional advantages of the VFN compared to classical coronagraphs include its relative insensitivity to telescope aperture shape, polarization aberrations, and many wavefront aberration modes \citep{Ruane2018_VFN}. Since its conceptual development, the VFN concept has been tested in the lab, achieving azimuthally averaged peak coupling of $16\%$ (close to the theoretical limit) and starlight suppression of $6 \times 10^{-5}$, which can be attributed to the minor wavefront errors in the system \citep[Monochromatic; Broadband,][]{Echeverri_VFN,echeverri_spie_2019}.

While the original VFN design is already compelling, it has several drawbacks. The planet throughput is relatively low, with a theoretical limit of $\sim 20\%$, depending on the configuration. The measurement from a VFN also lacks spatial information — since the coupling map is circularly symmetric, there is no way to determine from the data the position angle of the planet, information that is (in the absence of other measurements) necessary for constraining the orbital parameters of the planet. Since there is only one flux measurement and the coupling into the SMF varies with the radial separation of the planet, there is also a degeneracy between the planet flux and its separation. Here, we present an augmentation to the VFN that enhances throughput and provides additional constraints on the orbit and flux of the planet, while retaining the functionality of the VFN concept. This new design relies on a device called the mode-selective photonic lantern.

\subsection{Mode-Selective Photonic Lanterns} \label{sec:mspl}

A photonic lantern is a photonic mode converter that adiabatically interfaces between a multi-mode port and several single-mode ports, where the distribution of flux in the single-mode outputs is related to the power in each mode at the multi-mode input \citep{LeonSaval_PL_2013}. Photonic lanterns have been proposed for use in astrophysics for spectrometer coupling \citep{lin_2021} and for focal-plane wavefront sensing, allowing for the measurement of the input wavefront while maintaining single-mode fiber outputs suited for injection into spectrographs for spectral characterization \citep{jovanovic-ESS-2016,corrigan2018,Norris_2020}. Each mode at the few-mode fiber (FMF) face of the lantern is mapped to a SMF output, such that light coupling to a given mode at the FMF side will result in flux in the corresponding SMF core. The device is bi-directional, so light injected into one of the SMF ports will propagate into the mode corresponding to that port at the FMF face.

\begin{figure*}[!ht]
\begin{center}
    \includegraphics[width=0.4\textwidth]{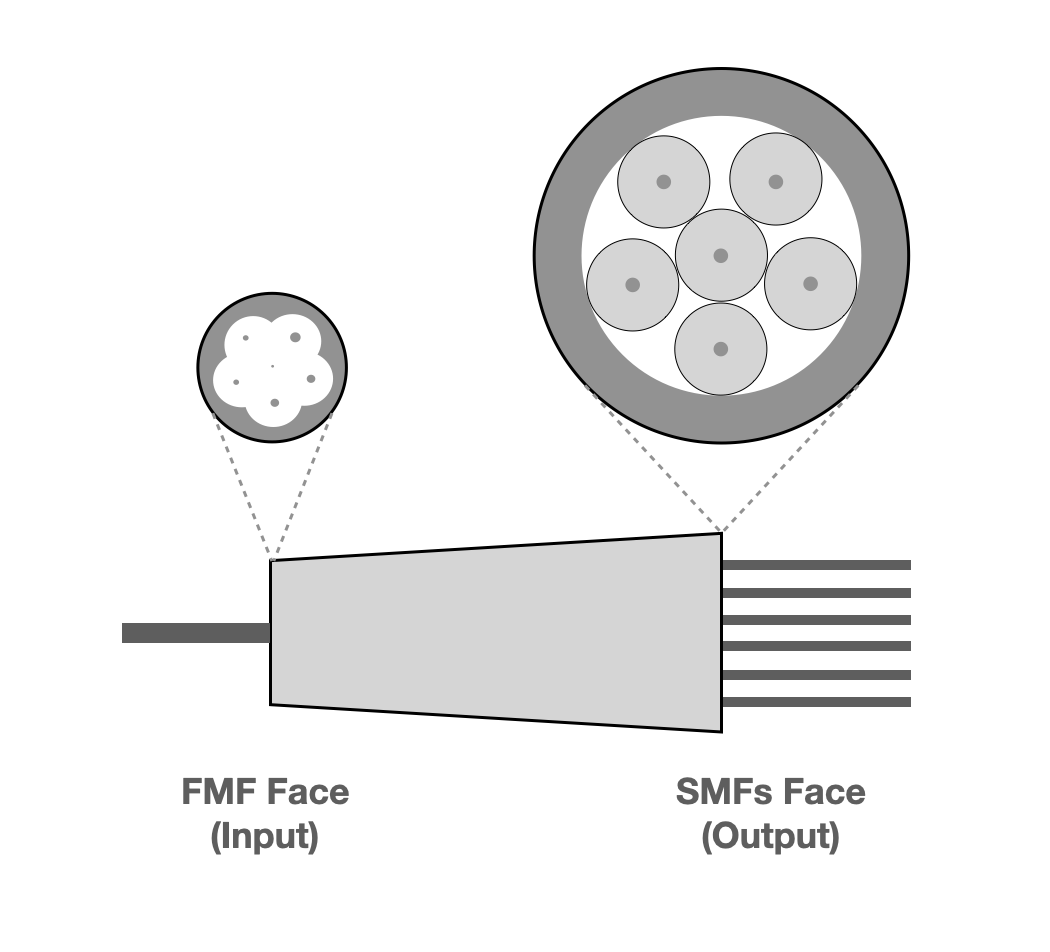}
    \includegraphics[scale = 0.65]{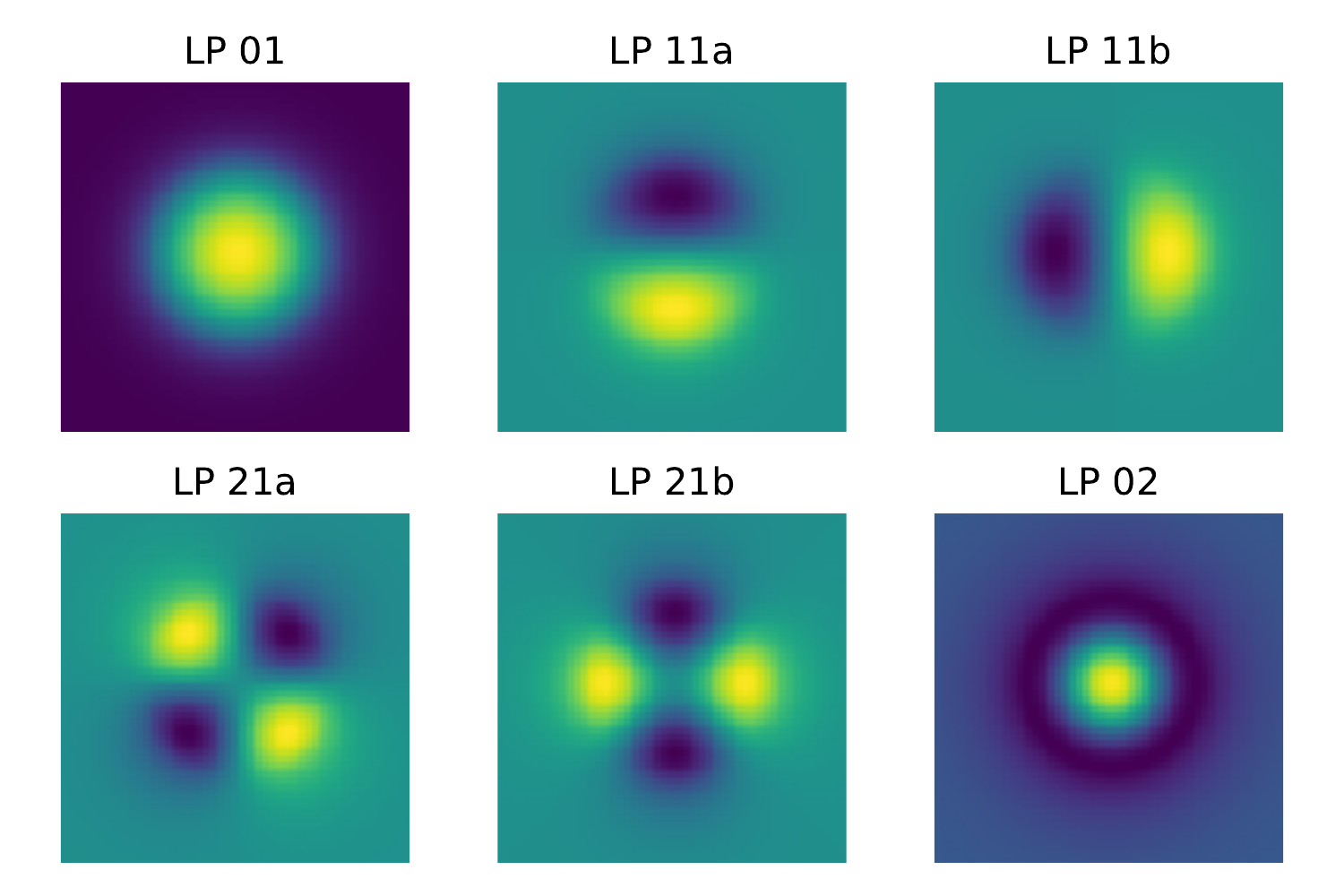}
    \caption{Left: Schematics of a 6-port mode-selective photonic lantern spatial-multiplexer fiber system. Each LP mode at the few mode fiber (FMF) face is mapped to one of the six single-mode ports of the SMF face, such that light with an LP mode shape at the FMF side will result in flux in the corresponding SMF core. The device is bi-directional, so light injected into one of the SMF ports will propagate into the LP mode corresponding to that port at the FMF face. Right: The field amplitudes of the first six LP modes, corresponding to the ideal modes of six-port MSPL.}
    \label{fig:mspl}
\end{center}
\end{figure*}

While standard photonic lanterns have similar cores and are not designed with a particular mode structure in mind, mode-selective photonic lanterns \citep[MSPL,][]{LeonSaval_MSPL} utilize dissimilar cores that enable ports to be mapped into LP modes, defined in \citet{lp_mode_def} as "the set of linearly polarized propagation modes of optical fibers with radially symmetric index profiles in the approximation of weak guidance." A partially mode-selective photonic lantern has one port corresponding to the LP 01 mode, while the rest of the ports exhibit an unspecified structure. In a fully mode-selective photonic lantern, all ports correspond to LP modes. Figure \ref{fig:mspl} shows a schematic of a six-port MSPL based on the design from \citet{LeonSaval_MSPL}, where each port corresponds to one of the first six LP modes.

To synergize the action of the VFN with symmetry properties of the LP modes, we propose to replace the single-mode fiber of the original VFN with a MSPL, resulting in a Photonic Lantern Nuller (PLN) instrument concept that improves upon the original design.

\subsection{VFN with a Mode-Selective Photonic Lantern}
The PLN replaces the single-mode fiber of the VFN by a MSPL as described in Section \ref{sec:mspl}. Specifically, the light after the vortex mask is focused onto the FMF face of the MSPL and propagates through to the single-mode outputs. Each output port can then be coupled into individual SMFs and routed to photodetectors or spectrographs. The port corresponding to the LP 01 mode provides the same response as the VFN, where on-axis light is nulled while off-axis light can couple. Additionally, if we label the LP mode azimuthal order by $m'$ analogously to the Zernike polynomials, i.e. positive $m'$ indicating an azimuthal component of $\cos(m'\theta)$ and negative $m'$ indicating $\sin(m'\theta)$, then, a photonic lantern port combined with an optical vortex with azimuthal charge $l$, will result in an on-axis null \textit{except} when $l\pm m'=0$. This result can be derived by extending Equation \ref{eq:overlap_integral} to an arbitrary fiber mode $\psi_{n'm'}$, and separating out the polar integral:

\begin{equation} \label{eq:lp_mode_overlap}
\begin{split}
    \int_0^{2\pi} \exp(il\theta) \cos(m'\theta) d\theta, \quad m' & \geq 0, \quad \mathrm{or}\\
    \int_0^{2\pi} \exp(il\theta) \sin(m'\theta) d\theta, \quad m' & < 0.
\end{split}
\end{equation}

Recalling the exponential trigonometric identities $\cos(x) = (e^{ix}+e^{-ix})/2$ and $\sin(x)=(e^{ix}-e^{-ix})/(2i)$, we find that these overlap integrals evaluate to 0 for $l \pm m' \neq 0$. Thus, on-axis nulls are created in multiple ports, from which planet spectra can be extracted. Additionally, the existence of ports with $m' \neq 0$ allows for a nuller configuration with no vortex at all, as the overlap integrals for the LP11ab and LP21ab ports evaluate to zero when $l=0$. This means that the photonic lantern can be used by itself as a nuller, as contemporaneously presented in \citet{Tuthill2022-NIH}.

To demonstrate these properties, we simulate the PLN configurations using HCIPy \citep{por2018hcipy}. Our optical propagation model propagates the desired input wavefront through a circular pupil (with $\lambda/D$ chosen to equal 1), and then into a focal plane. For the configuration without a vortex, this becomes the final focal-plane electric field. For the configurations with a vortex, either a charge 1 or 2 vortex is applied in the focal plane. As with the VFN, a vortex with charge higher than 2 results in lower peak throughput and larger IWA, so we do not focus on them in this work.

The square of the overlap integral of the focal-plane electric field distribution with each LP mode gives the relative intensity coupled into the corresponding port. We explore using an MSPL with six LP modes and a V number, "the normalized frequency parameter that determines the number of modes" \citep{vnum_def}, equal to 4.71. Our simulations assume perfect mode shapes as well as perfect transitions, free from cross-coupling and losses. Characterizing the impact of these real-world imperfections, from realistic designs as well as from fabrication errors, is left for future work.

Given wavelength, the optimal coupling into the lantern depends on the the mode field diameter (MFD) of the lantern modes and the focal ratio $F\#$ \citep{ruane_2019_spie}. While the real MFDs of photonic lanterns are tunable within a small range (Leon-Saval, private communication), in practice, the coupling in a real system will be optimized by changing the focal ratio. However, since our simulations already set $\lambda/D=1$ and $F\#=1$, we optimize coupling by tuning the MFD (expressed in units of $\lambda/D$). Specifically, for each configuration (no vortex, $l=1$, $l=2$), we simulate a range of MFDs and find the value that maximizes the peak of the x-axis cross-section of the summed throughput of the nulled ports. Although Section \ref{ssec:detection} shows that summed throughput does not fully predict instrument performance, it is still a useful proxy for choosing the MFD, as optimizing directly for detection capability would require knowledge of the level and distribution of on-sky wavefront error, which is not predictable \textit{a priori}.

\begin{figure*}[t]
\begin{center}
    No Vortex \hspace{120pt} Charge 1 \hspace{120pt} Charge 2
    \fbox{\includegraphics[scale = 0.35]{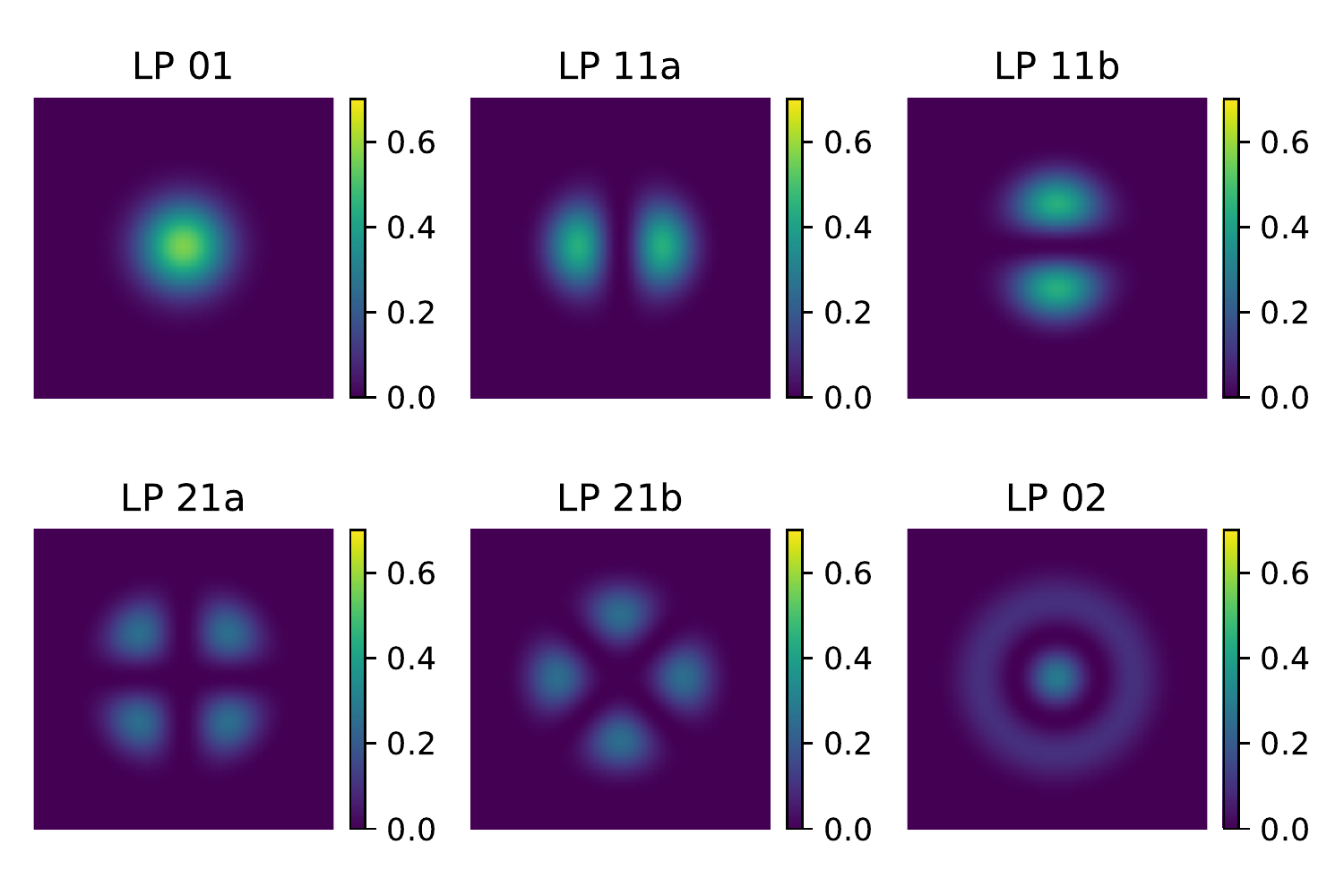}}
	\fbox{\includegraphics[scale = 0.35]{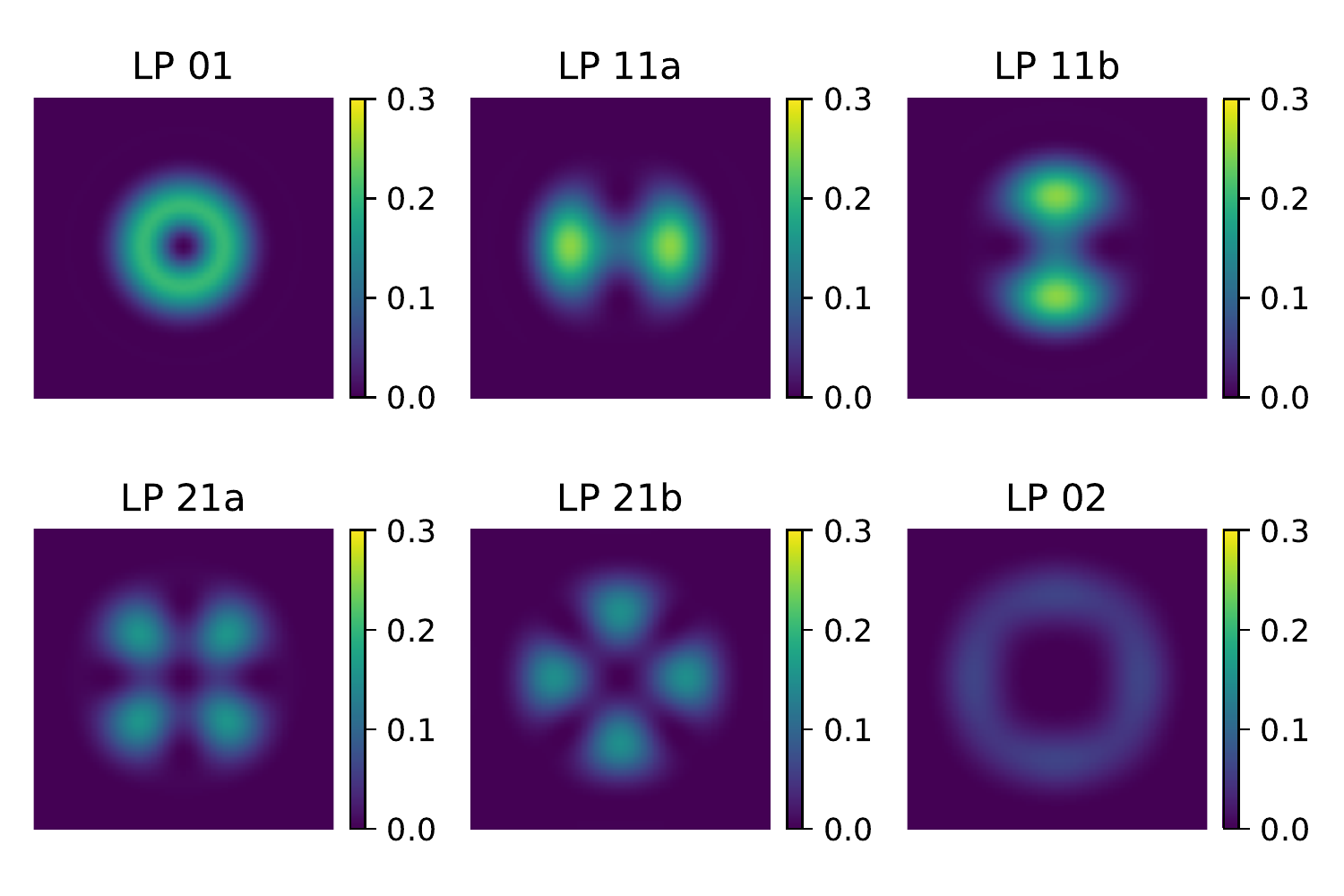}}
	\fbox{\includegraphics[scale = 0.35]{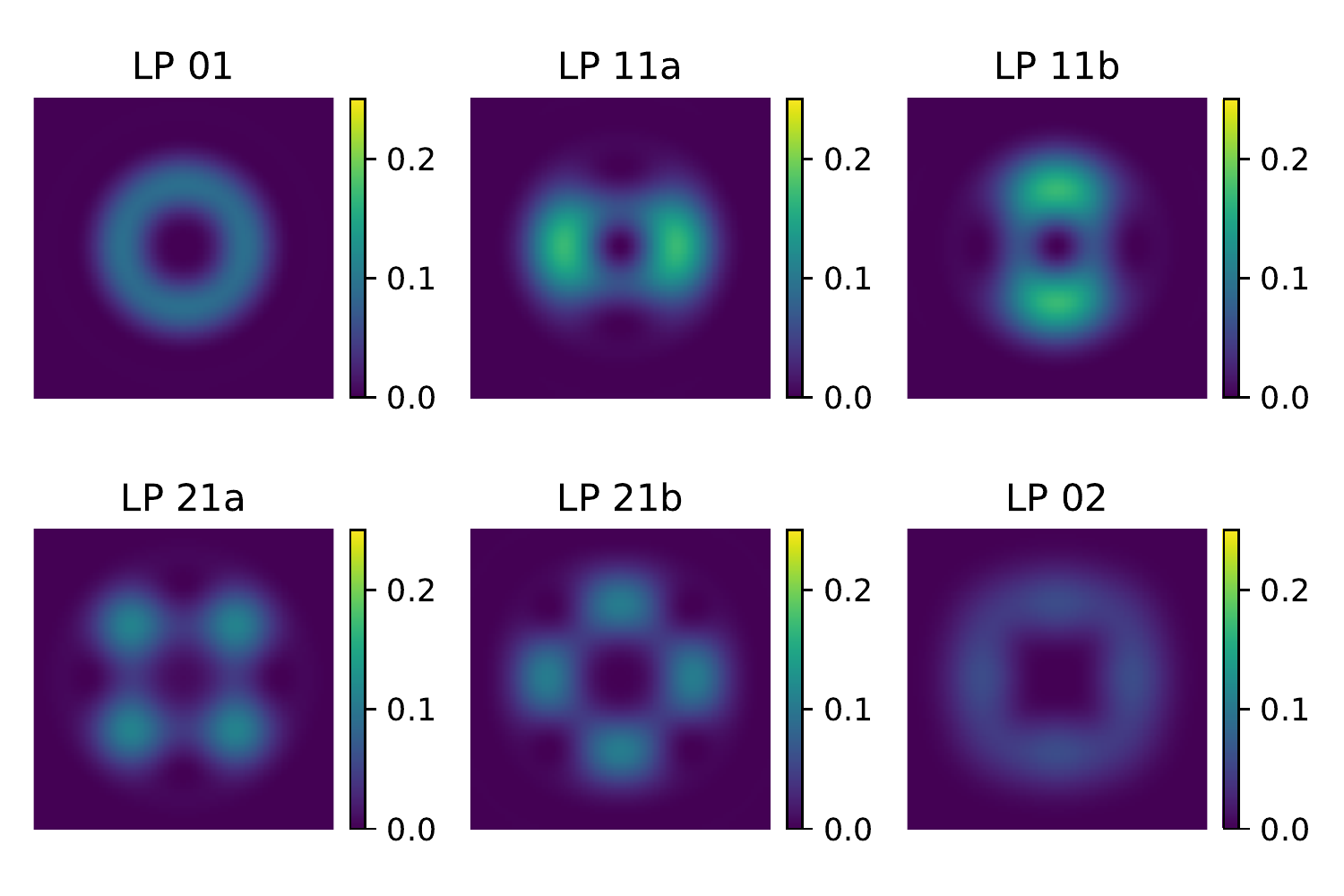}}
	\includegraphics[scale = 0.37]{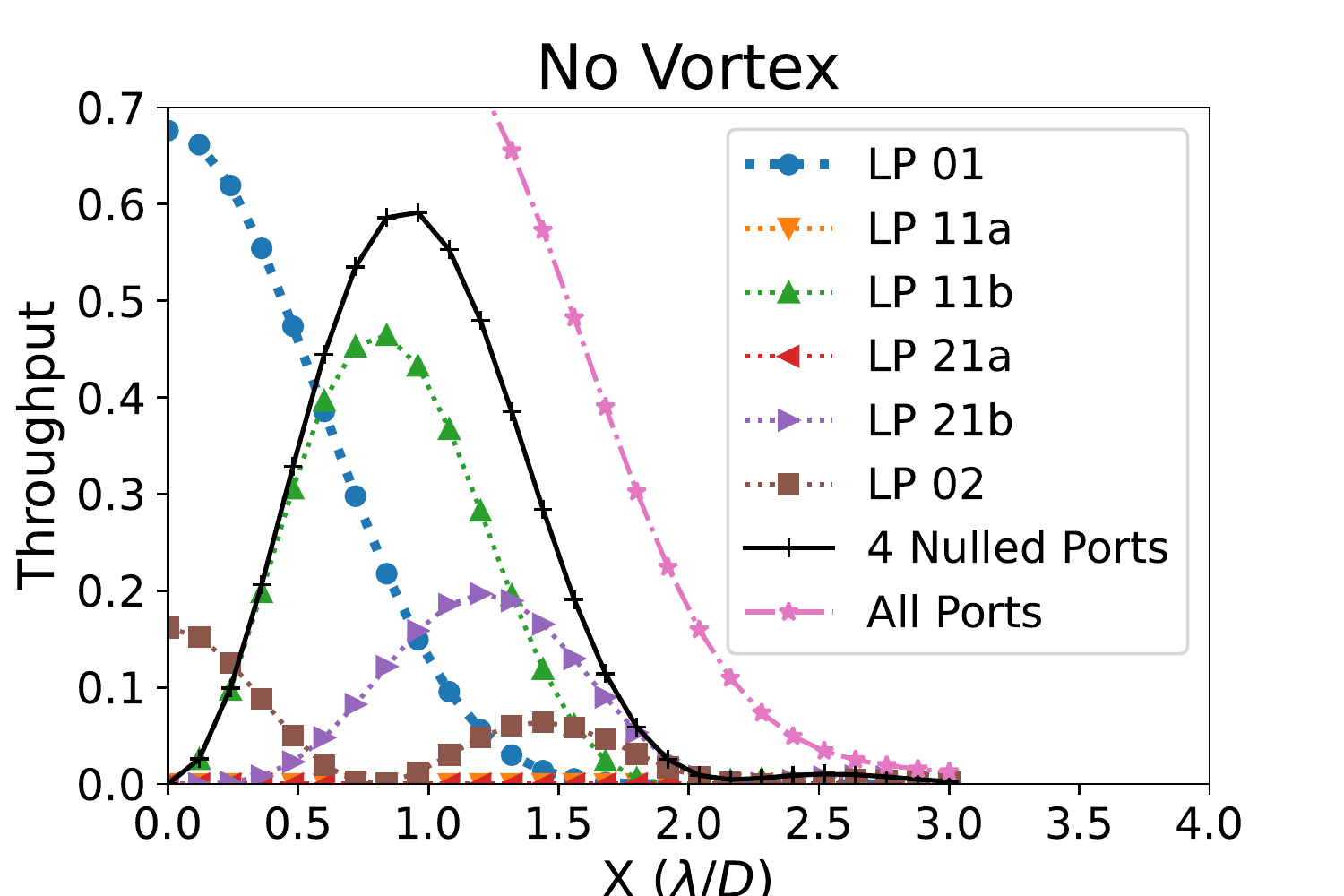}
	\includegraphics[scale = 0.37]{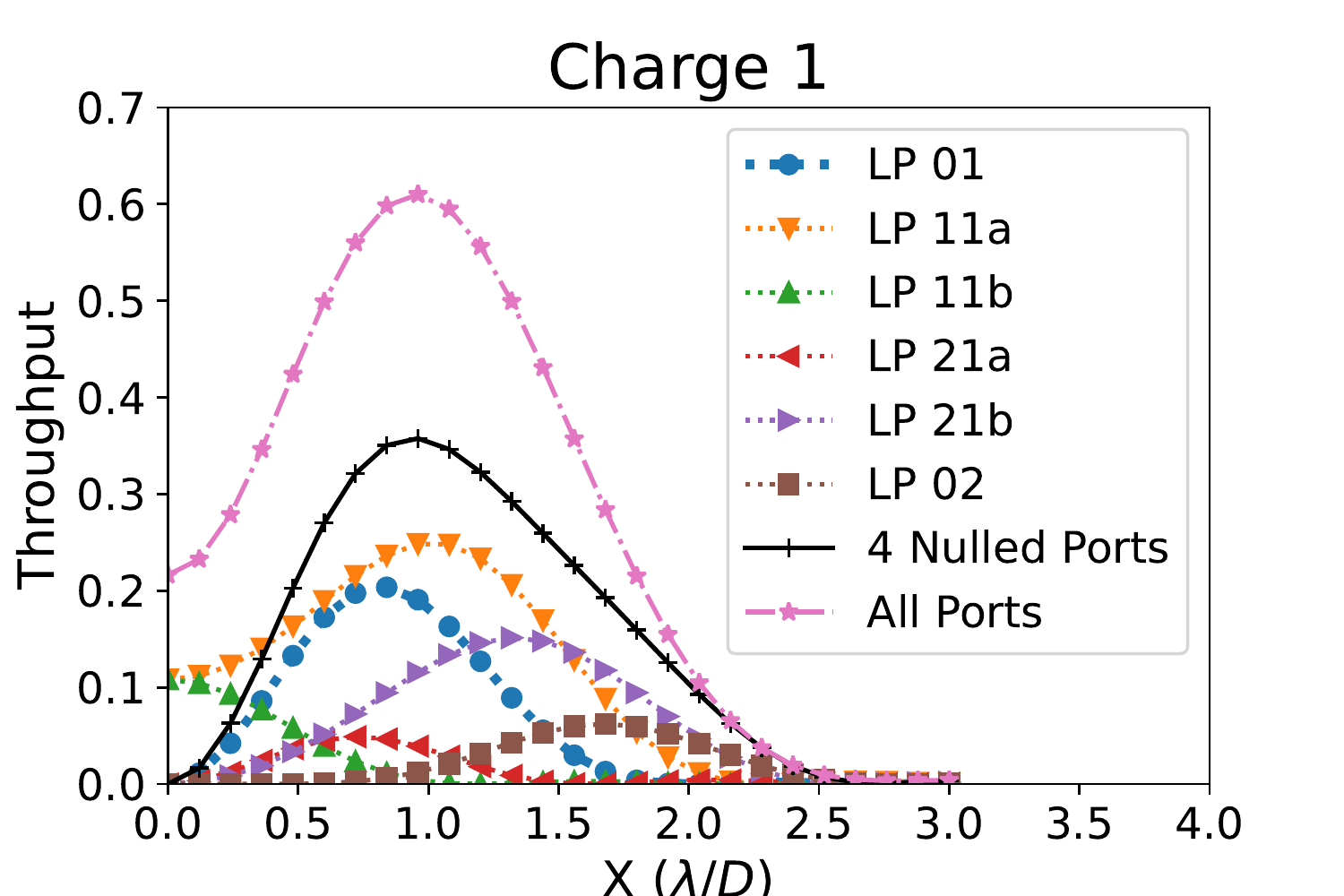}
	\includegraphics[scale = 0.37]{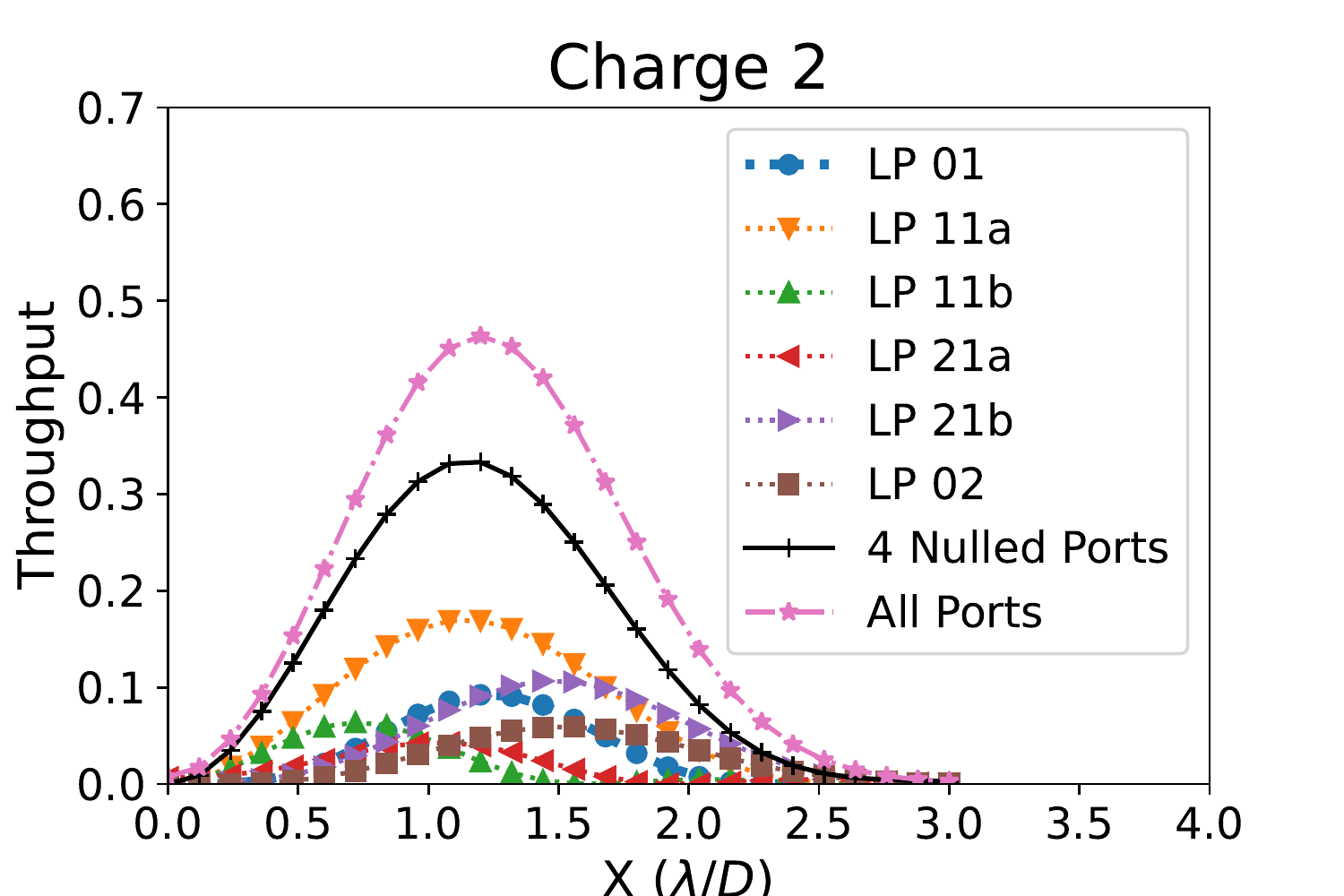}
	\caption{\label{fig:coupling_donuts} Coupling maps for each port with no vortex (top left), and a charge 1 (top middle) and charge 2 (top right) vortex. The maps span -3 $\lambda/D$ to 3 $\lambda/D$ in each direction. Bottom left: Throughput line profiles with no vortex. The four nulled ports satisfying $l\pm m'=0$ are LP 11ab and LP 21ab. Bottom middle: Throughput line profiles with a charge 1 vortex. The four nulled ports satisfying $l\pm m'=0$ are LP 01, LP 21ab, and LP02. Bottom right: Throughput line profiles for each port with a charge 2 vortex. The four nulled ports satisfying $l\pm m'=0$ are LP 01, LP 11ab, and LP02. Although nulls in the LP 21ab ports are not guaranteed by symmetry, in this case, their central throughputs are spuriously low, and including them in the data analysis may provide some additional gains. 
	}
\end{center}
\end{figure*}

From our simulations, we find that the optimal MFD is 2.8 $\lambda/D$ for the no vortex and charge 1 cases, and 3.2 $\lambda/D$ for the charge 2 case. We present the results of our simulations using these diameters. Figure \ref{fig:coupling_donuts} shows the ideal spatial coupling efficiency for a point source as a function of angular separation from the optical axis, or coupling map, for every port (top panels) along with the line profile along the horizontal axis (bottom panels). We also plot the total flux collected across all ports (dashed pink lines) as well as the total flux collected from only the nulled ports satisfying $l\pm m'=0$ (solid black lines). The total nulled throughput curves demonstrate that the additional ports increase both the peak throughput as well as the field of view for which planet light couples.

\begin{figure*}[t]
\begin{center}
    \includegraphics[scale = 0.38]{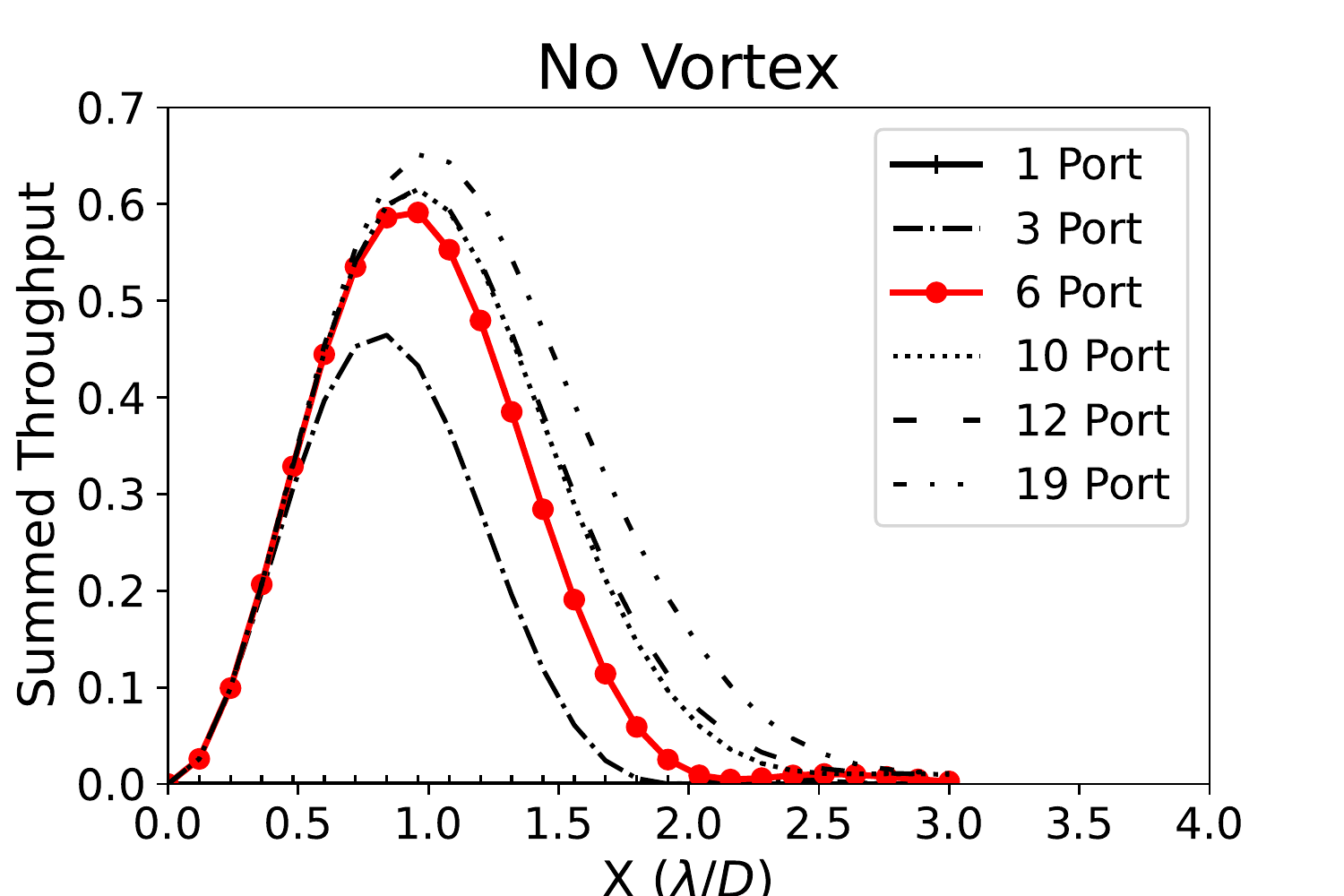}
	\includegraphics[scale = 0.38]{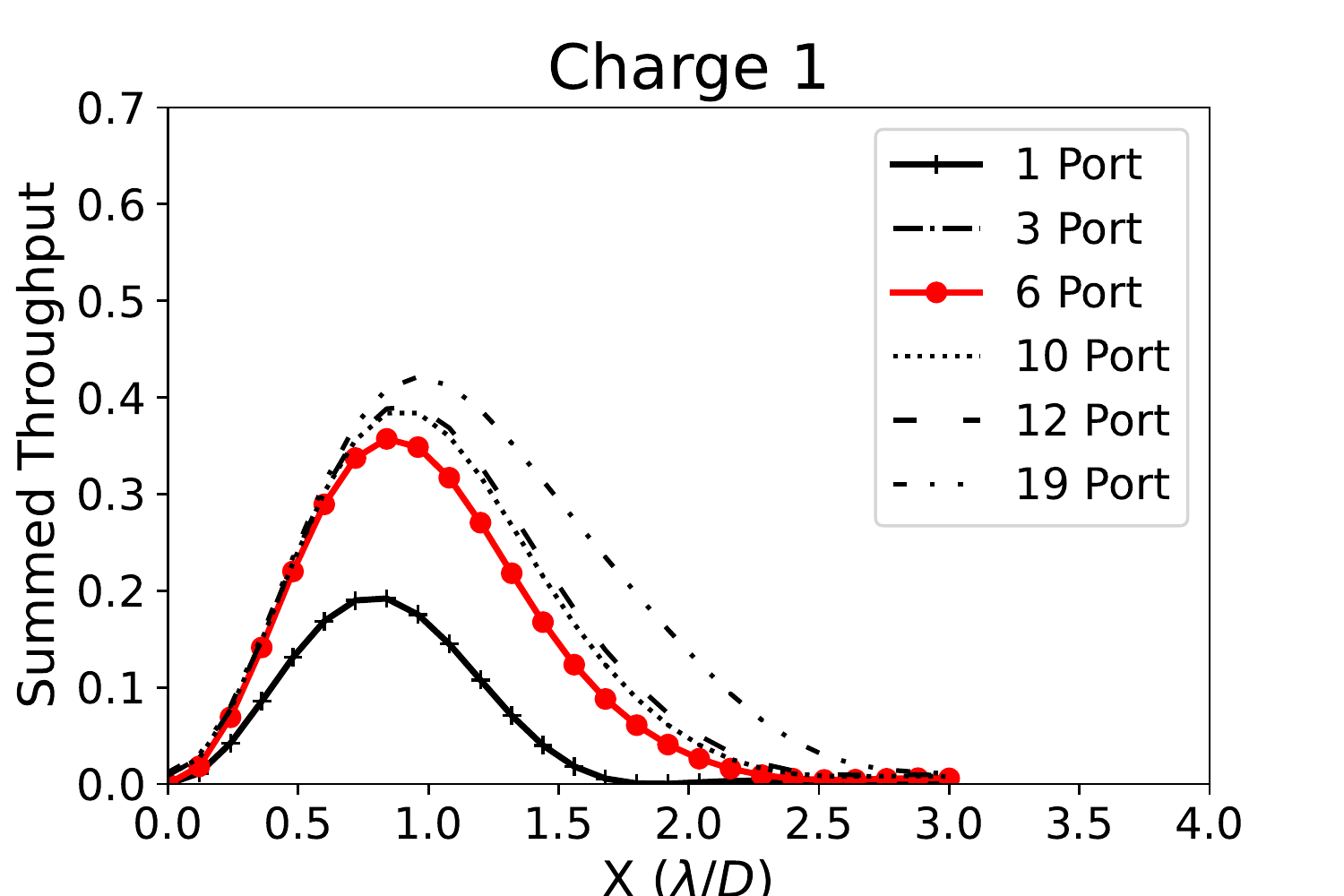}
	\includegraphics[scale = 0.38]{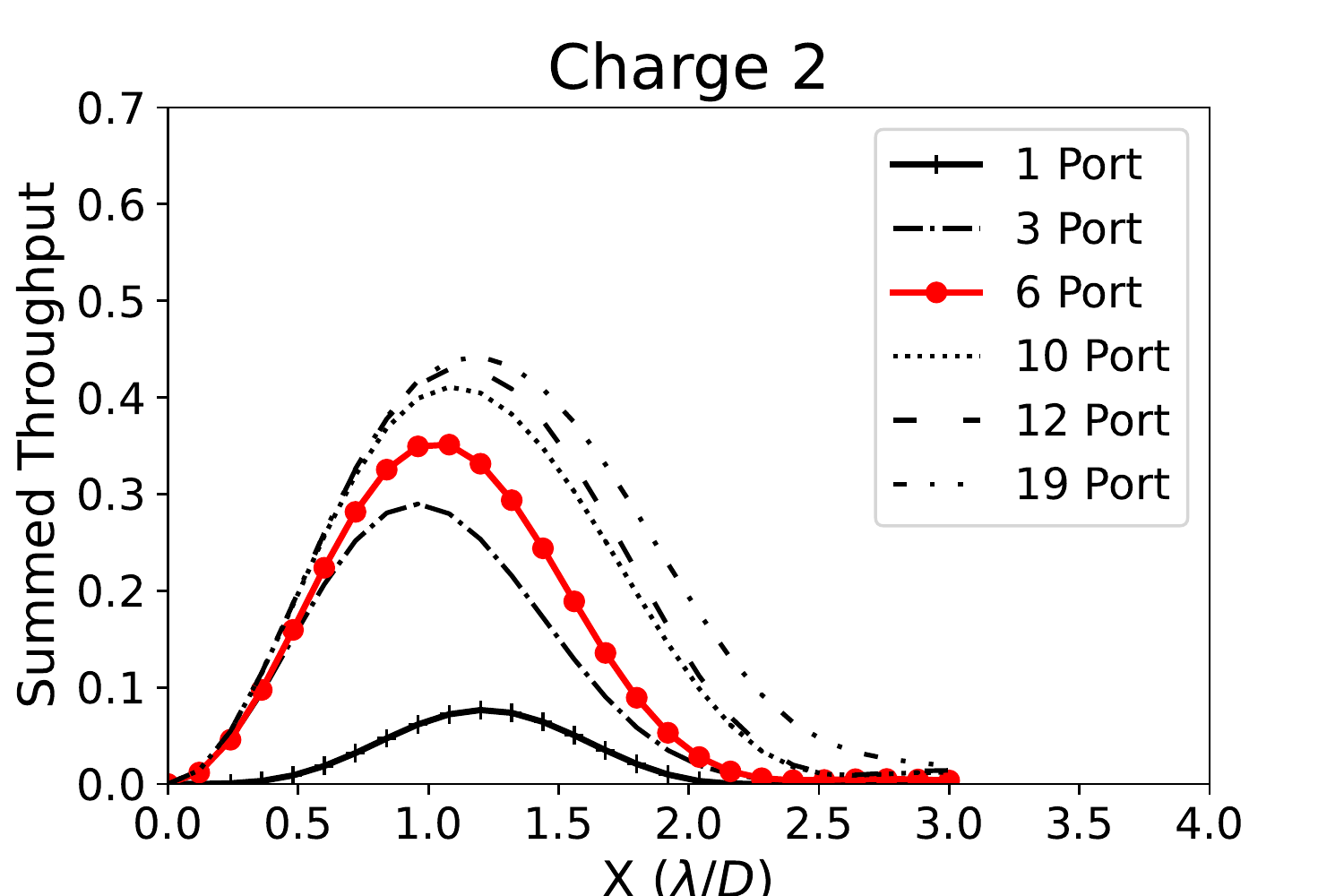}
	\caption{\label{fig:port_num_coupling} Line profiles for summed throughput of nulled ports for PLNs with no vortex (left), a charge 1 vortex (middle) and a charge 2 vortex (right), using MSPLs with varying numbers of output ports. As the number of ports increases, each additional port brings decreasing returns in additional throughput. The current limit of what can be practically manufactured is six ports. Thus, we choose to use a six-port MSPL in our PLN design, which balances the total throughput of the nulled ports with what is practically manufacturable. Note that a higher V number of 8.48 was necessary to generate up to 19 LP modes. Here, we wish to compare the effect of port number independently of V number effects, so fix the V number at 8.48 for all port numbers. Thus, due to the difference in V number, the line profiles shown in this analysis have slightly different shapes from those in Figure \ref{fig:coupling_donuts}.}
\end{center}
\end{figure*}

While MSPLs with more than six ports can in theory be fabricated, manufacturing MSPLs with large numbers of modes remains a practical challenge because the adiabaticity of the lantern transition becomes more difficult to achieve as the number of modes increases \citep{Velazquez-Benitez2018}. While larger port numbers may become available with the advancement of photonics technology, Figure \ref{fig:port_num_coupling} shows that increasing the total number of ports brings diminishing returns in throughput, especially at angular separations $<\lambda/D$. In addition, using fewer ports has the advantage that it requires fewer detector pixels, which are always at a cost premium. Considering these factors, and that MSPLs with more than six ports are not readily manufacturable with current photonics technology, we choose to focus our investigations on a PLN design with a six-port MSPL.

\section{Sensitivity to Aberrations} \label{sec:aberrations}

\subsection{Zernike Aberrations}

One benefit of the original VFN was its insensitivity to many low order Zernike wavefront error modes. If the charge of the vortex is denoted by $l$, and the Zernike aberrations are denoted by $Z_n^m(r,\theta)$, where $n$ is the radial order and $m$ indicating the azimuthal structure, i.e. $\cos(m\theta)$ for positive $m$ and $\sin(m\theta)$ for negative $m$, then only aberrations that cancel out the vortex charge ($l\pm m=0$) will couple. This can be demonstrated analogously to the case of LP modes, replacing the $m'$ of a given port in Equation \ref{eq:lp_mode_overlap} with the $m$ of a given Zernike mode.

\begin{figure*}[t]
\begin{center}
    No Vortex
    
	\includegraphics[scale = 0.28]{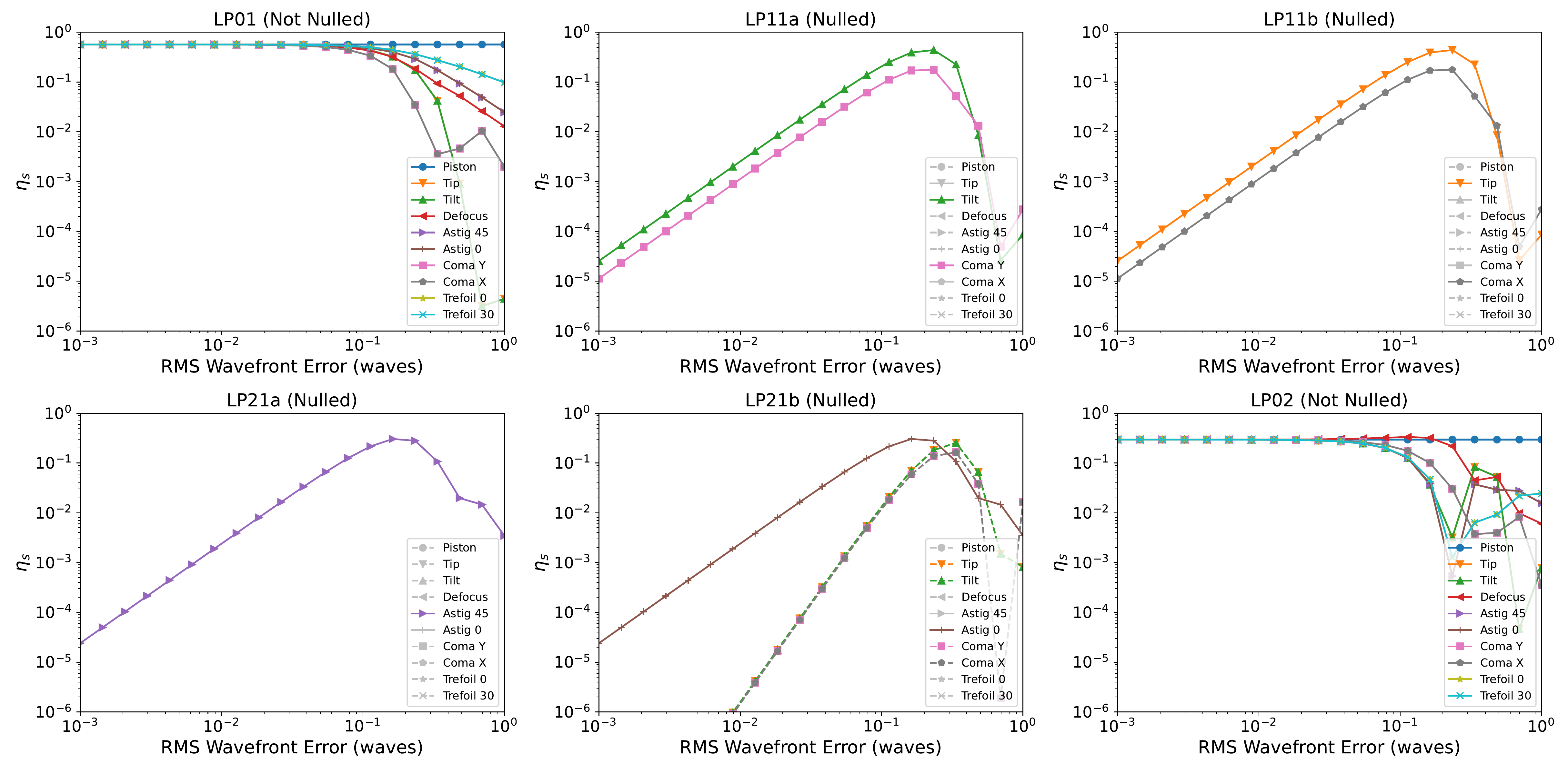}

    Charge 1
    
	\includegraphics[scale = 0.28]{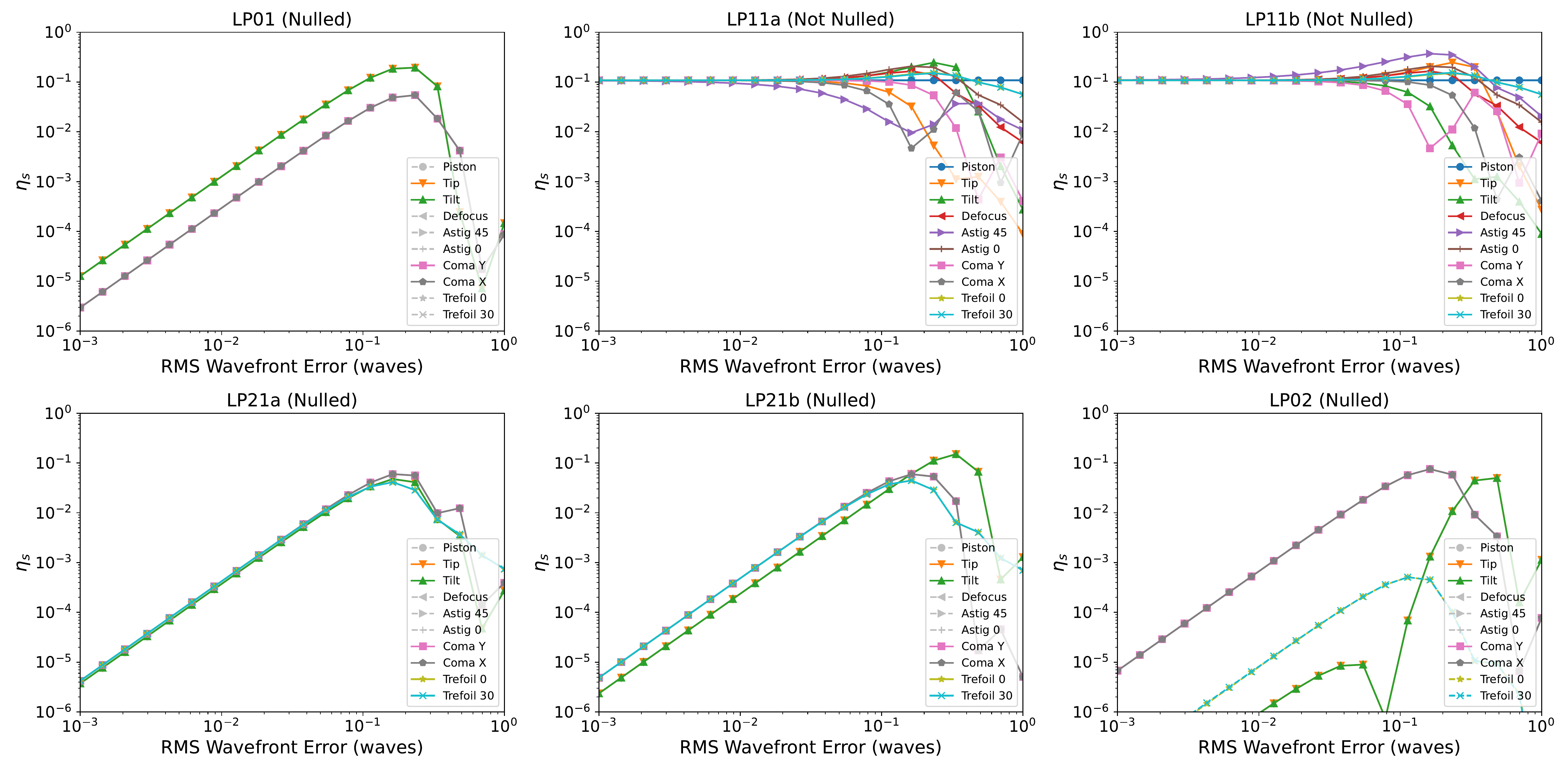}
	
	Charge 2
	
	\includegraphics[scale = 0.28]{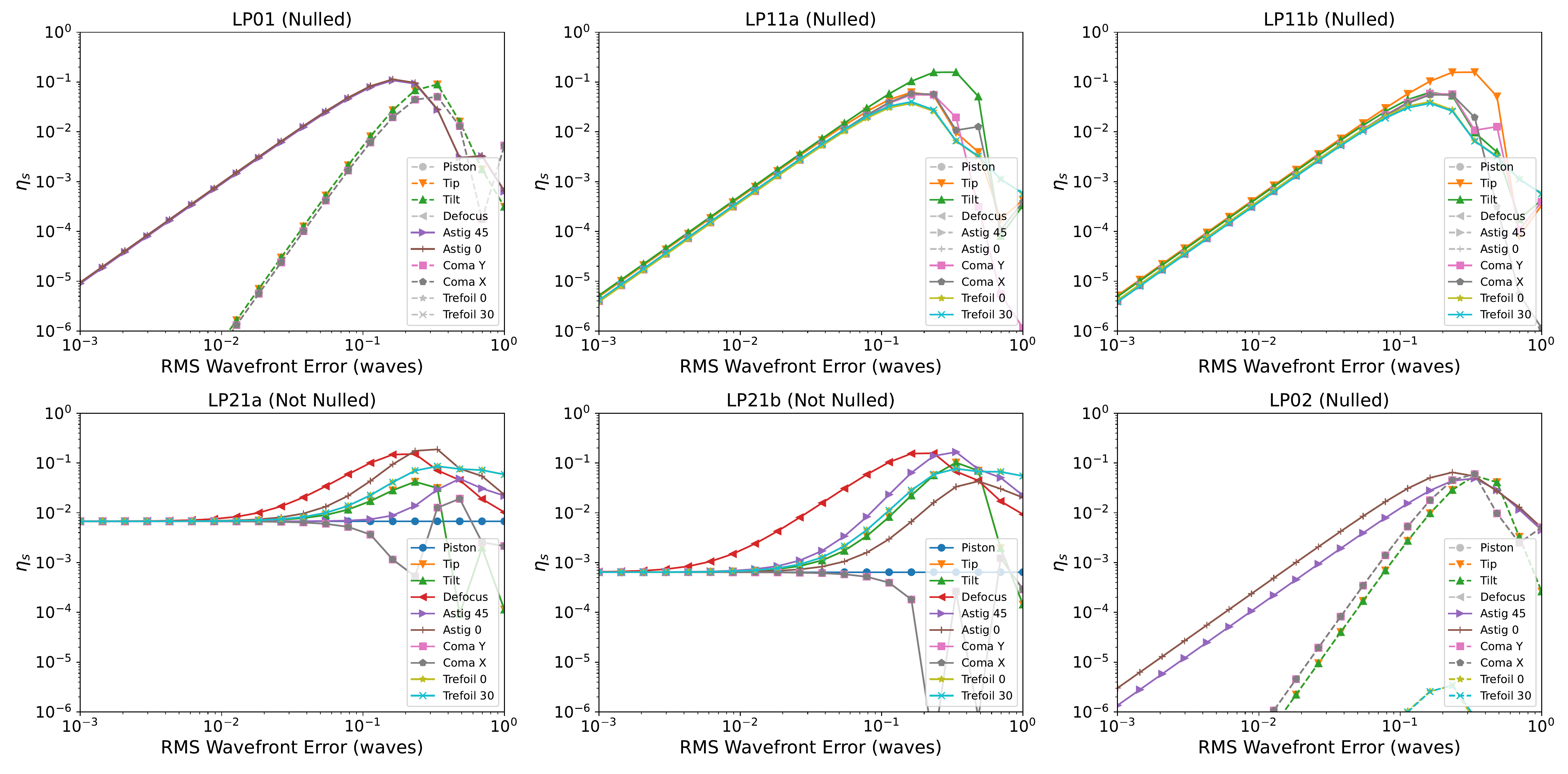}
	\caption{\label{fig:z_sensitivities} Stellar coupling rate as a function of individual Zernike polynomial amplitude, with no vortex (top), a charge 1 vortex (middle), and a charge 2 vortex (bottom). For the nulled ports, solid lines indicate modes predicted to couple (those satisfying $l \pm (m' + m) =0$), while dashed lines indicate modes that are not predicted to couple (to first order, though higher-order coupling effects can be seen). Values of $\eta_s$ falling below $10^{-6}$ are likely numerical noise, and are not shown. Lines that fall entirely below $10^{-6}$ are light grey in the legend.}
\end{center}
\end{figure*}

The additional photonic lantern ports obey a similar principle, but the structure of the LP mode and the Zernike mode will interact, and the polar overlap integral is now given by

\begin{equation} \label{eq:full_overlap_integral}
\begin{split}
    & \int_0^{2\pi} \exp(il\theta) \cos(m'\theta) \cos(m\theta) d\theta, \quad m',m \geq 0, \quad \mathrm{or}\\
    & \int_0^{2\pi} \exp(il\theta) \cos(m'\theta) \sin(m\theta) d\theta, \quad m' \geq 0, m < 0, \quad \mathrm{or}\\
    & \int_0^{2\pi} \exp(il\theta) \sin(m'\theta) \cos(m\theta) d\theta, \quad m' < 0, m \geq 0, \quad \mathrm{or}\\
    & \int_0^{2\pi} \exp(il\theta) \sin(m'\theta) \sin(m\theta) d\theta, \quad m',m < 0.
\end{split}
\end{equation}

Thus, for each port, only aberrations satisfying $l \pm (m' + m) =0$ will couple (to first order). Figure \ref{fig:z_sensitivities} shows the simulated stellar coupling, $\eta_s$, as a function of the input amplitude of the first ten Zernike aberrations. In this work, we compute coupling normalized to the summed intensity of the beam, such that the stellar coupling is equivalent to the null-depth. The fact that the LP 01 port is sensitive primarily to tip, tilt, and coma (for charge 1) and astigmatism followed by second-order responses to tip and tilt (for charge 2) is consistent with theoretical predictions as well as the numerical simulations presented in \citet{ruane_2019_spie}. The results for the other ports show that, as predicted by the azimuthal order conditions, each port is only sensitive to a few specific lower-order aberrations satisfying $l \pm (m' + m) =0$. For example, the LP 21ab ports with a charge 1 vortex and the LP 11ab ports with a charge 2 vortex are all insensitive to defocus ($m=0$) and astigmatism ($m=\pm2$). The LP 02 ports have the same azimuthal order as the corresponding LP 01 ports, and thus reject the same low-order aberrations.

\subsection{Tip-tilt Jitter}

\begin{figure*}[t]
\begin{center}
    \includegraphics[scale = 0.38]{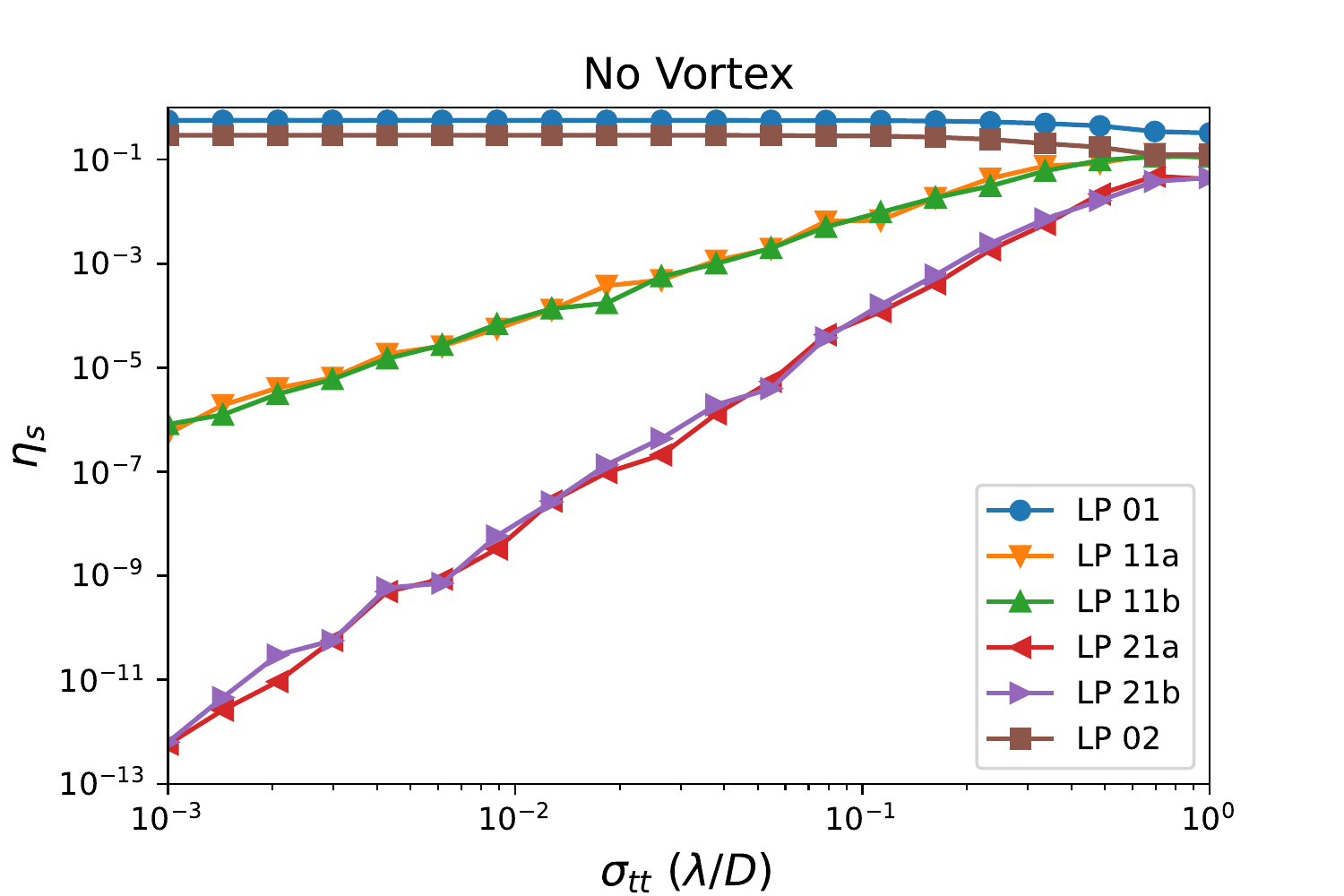}
	\includegraphics[scale = 0.38]{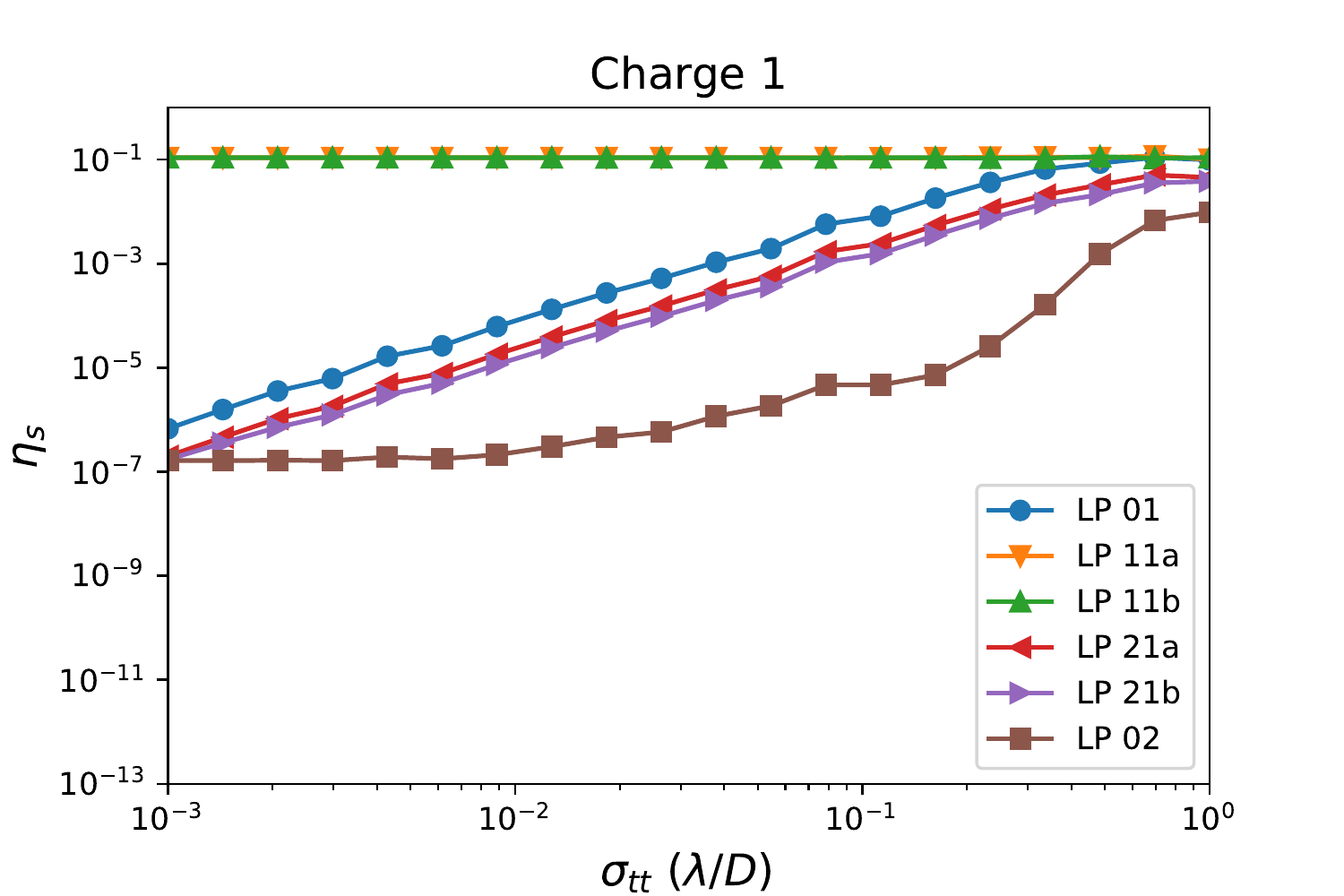}
	\includegraphics[scale = 0.38]{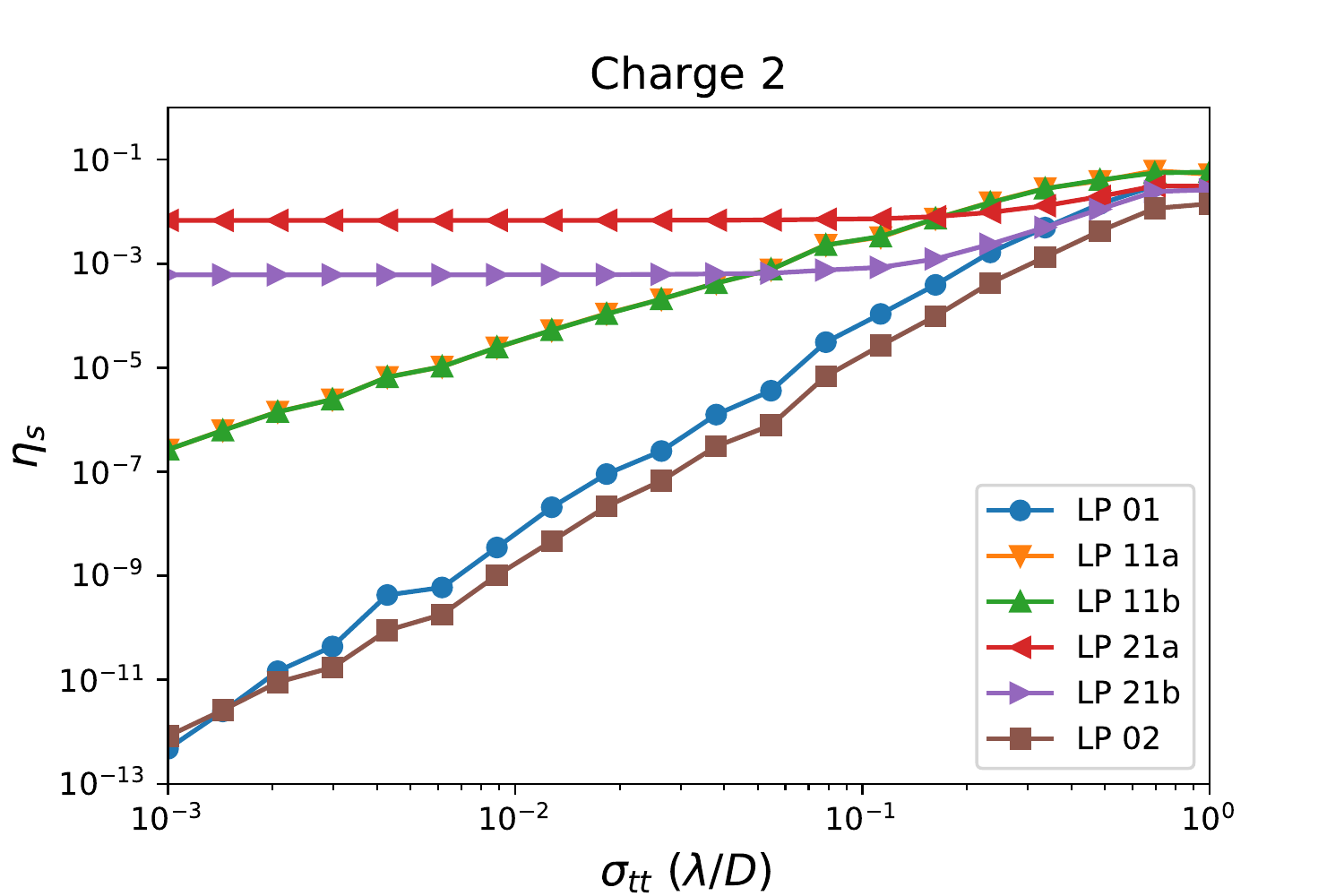}
	\caption{\label{fig:jitter_coupling} Left: Stellar coupling rates as a function of tip-tilt jitter, random-uniformly distributed in position angle, with no vortex (left), a charge 1 vortex (middle), and a charge 2 vortex (right). The standard deviation of the per-frame tip-tilt amplitude is given by $\sigma_{tt}$, with position angle drawn uniformly between 0 and $2 \pi$.}
\end{center}
\end{figure*}

\citet{ruane_2019_spie} predicted that for ground-based observatories, tip-tilt jitter (evolving much faster than the typical exposure times) will likely be a significant contribution to degradation of the VFN's null-depth. We thus present simulations of average null-depth achieved ($\eta_s$) as a function of the standard deviation of tip-tilt jitter ($\sigma_{tt}$). For each data point, 100 independent realizations of tip-tilt are generated, with amplitude drawn from a normal distribution with standard deviation $\sigma_{tt}$ and position angle drawn uniformly between 0 and $2 \pi$. The 100 frames are then averaged to calculate an averaged $\eta_s$. The results are presented in Figure \ref{fig:jitter_coupling}. For example, to achieve a null depth of $10^{-3}$ in the LP11ab ports of the no vortex PLN, the standard deviation of tip-tilt jitter must be smaller than $\sim 0.1 \lambda/D$. To achieve a null depth of $10^{-3}$ in the LP01 port of the charge 1 and charge 2 configurations, the standard deviation of tip-tilt jitter must be smaller than $\sim 0.1 \lambda/D$ and $\sim 0.3 \lambda/D$, respectively. For context, the Keck Planet Imager and Characterizer (KPIC) instrument at the Keck II telescope, a fiber injection unit for high resolution spectroscopy that currently has an VFN mode as well as the capability to test a future PLN on-sky, typically achieves on-sky jitter standard deviations of 6-7 mas, corresponding to 0.14 waves at 2.2 $\mu$m \citep{delorme_2021}.

\subsection{KPIC Atmospheric Residuals}

We also simulate the performance of the PLN under WFE conditions measured by the pyramid wavefront sensor (PyWFS) of KPIC. The atmospheric seeing the night the data was taken was 0.6 arcsec, and the wavefront sensor achieved residuals of 150 nm RMS. It should be noted that the PyWFS does not see all of the errors in the optical system, as recent on-sky demonstrations of the VFN on KPIC (Echeverri et al, in prep) do not achieve the level of starlight suppression predicted by these residuals alone. Specifically, in the real KPIC instrument, there is additional tip-tilt error downstream of the PyWFS that is not captured in these simulations. Thus, these simulations should be interpreted as an optimistic limit, while the real performance will be impacted by additional errors invisible to the PyWFS.

\begin{figure*}[!ht]
\begin{center}
	\includegraphics[scale = 1.0]{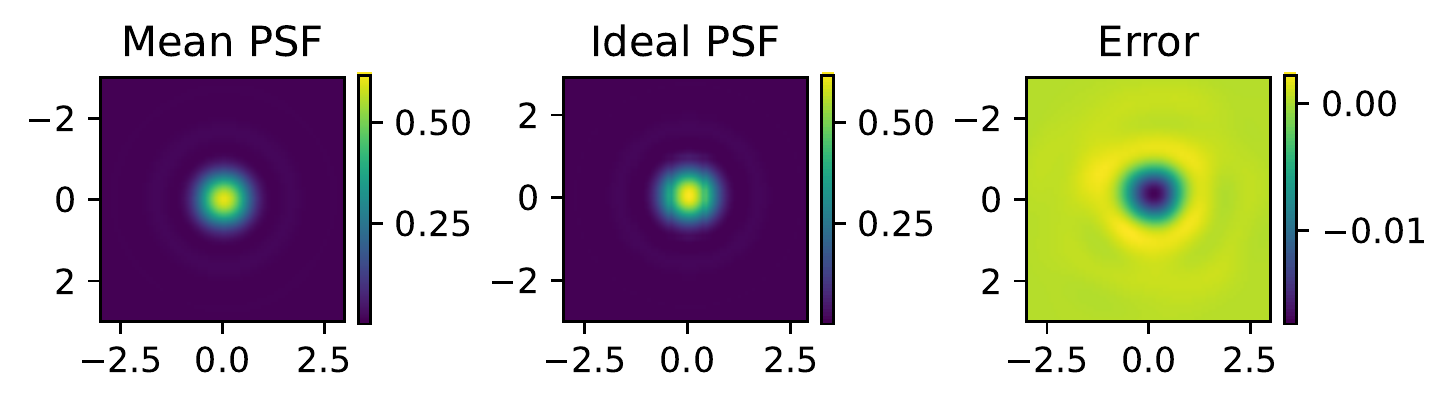}
	\caption{\label{fig:psfs} Left: Mean focal-plane PSF in the presence of WFE as measured by the KPIC PyWFS. Middle: Unaberrated focal-plane PSF. Right: Difference between the aberrated and ideal PSFs. Reminder that these are \textbf{not} simulations of the Keck PSF, but of the measured wavefront error residuals propagated through a system with an ideal circular aperture.}
\end{center}
\end{figure*}

For our simulation, we take 590 frames of measured wavefront error, expressed in the form of reconstructed Zernike coefficients. From each frame of coefficients, we generate a pupil plane WFE map. As an intermediate diagnostic, we calculate the focal-plane image PSF averaged over these frames, compared it to an ideal PSF with no WFE in Figure \ref{fig:psfs}.

For our simulation, we propagate an on-axis beam with that WFE through our PLN models to calculate the output null depths. We also propagate off-axis beams with each frame of WFE (at $0.84 \lambda/D$ for no vortex and charge 1 configurations and $1.3 \lambda/D$ for charge 2 configuration). The instantaneous coupling over time with these residuals may be found in Appendix \ref{app:timeseries}. Meanwhile, Figure \ref{fig:kpic_coupling} shows the mean coupling over all the frames. In the nulled ports of the PLN, the mean off-axis planet coupling over these frames (where it is expected based on the coupling maps) remains significantly higher than the stellar coupling in the presence of this WFE.

\begin{figure*}[!ht]
\begin{center}
	\includegraphics[scale = 0.45]{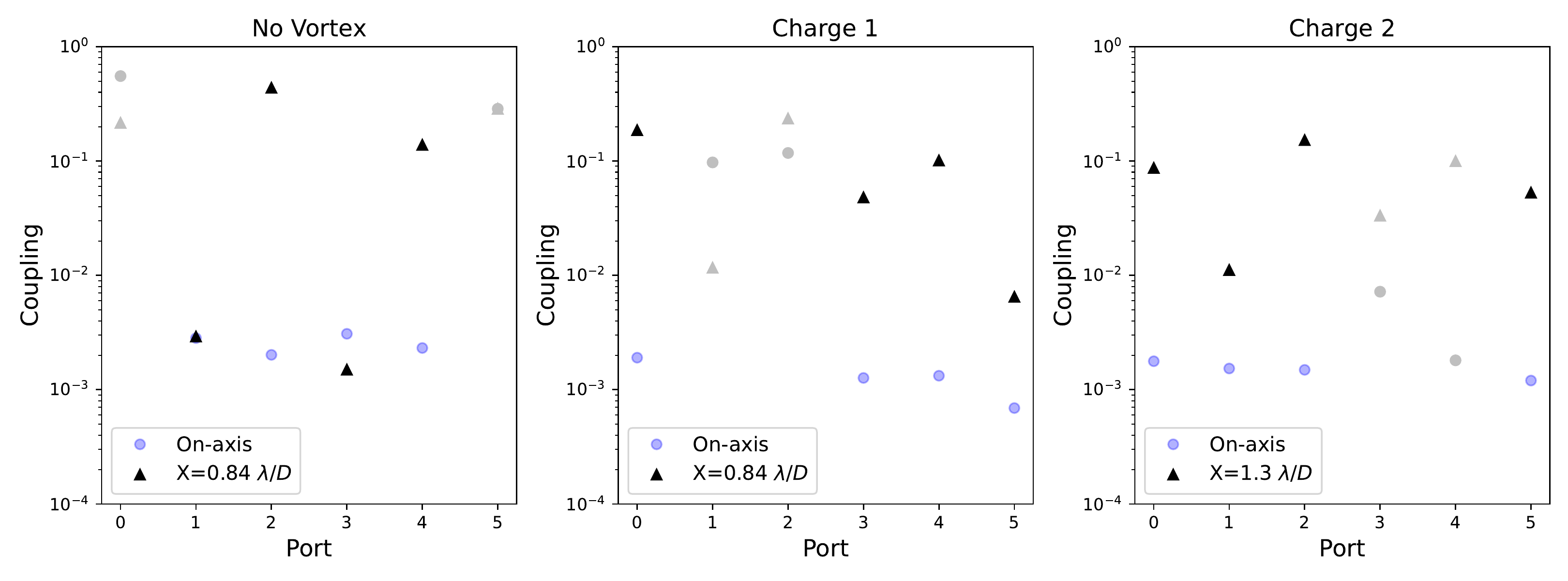}
	\caption{\label{fig:kpic_coupling} Mean coupling calculated over 590 frames of WFE residuals from the KPIC PyWFS, for the no vortex (left), charge 1 (middle), and charge 2 (right) configurations. The ports on the bottom axis are (from left to right): LP01, LP11a, LP11b, LP21a, LP21b, and LP02. Coupling values for ports that are not considered nulled are depicted in light grey. Off-axis planet coupling (where it is expected based on the coupling maps) remains higher than the stellar coupling in the presence of these WFE realizations.}
\end{center}
\end{figure*}

\section{Simulation of Exoplanet Characterization}

In this section, we demonstrate the exoplanet detection and characterization capabilities of a PLN and compare it to those of the VFN. 

\subsection{Synthetic Data Generation} \label{sec:datagen}

We consider the outputs of the instrument to be the intensity at the single simulated wavelength in each port. In reality, the light in each port can be fed in to a spectrograph, and spectral analysis can be used to increase detectability by orders of magnitude \citep{wang_ji_2017}. However, we neglect spectral information in this preliminary demonstration of the PLN performance relative to the VFN, and leave exploring the combination of a broadband PLN and spectral analysis to future work.

We assume that the integration time of an observation is significantly longer than the coherence time of atmospheric residuals, such that fluctuations in wavefront error will average out to the null depth. Consequently, we assume that the primary contribution to non-static noise is photon noise.

The following process was used to generate the synthetic data. We first average the 590 intensity frames from the simulation of KPIC PyWFS residuals in Section \ref{sec:aberrations} to obtain the average null depth. To generate realizations of photon noise, we calculate the stellar photon rate entering the instrument:

\begin{equation} \label{eq:pr}
    \mathrm{PR} = f_0 \times 10^{-m/2.5} \times A \times \Delta \lambda \times \eta_t,
\end{equation}

where $f_0 = 9.56\times 10^9$ photons m$^{-2}$ s$^{-1}$ $\mu$m$^{-1}$ is the zero point number corresponding to the photon flux per unit wavelength of a magnitude zero star in H band, $m$ is the stellar magnitude, $A$ the telescope area, $\Delta \lambda$ the bandwidth, and $\eta_t$ the throughput of the telescope before reaching the PLN instrument. We choose the stellar magnitude to be $m=5$ and use the Keck telescope area ($A=76$ m$^2$). We assume a bandwidth of $\Delta \lambda = 0.15 \mu$m and upstream telescope throughput of $\eta_t = 0.06$, a typical value for Keck.

For each port of the PLN, we multiply PR by its null depth to calculate the photon rate per port. We then multiply that photon rate by the assumed exposure time of 60 s to obtain the counts per exposure. We add normally-distributed noise with a variance equal to the number of counts, an approximation for Poisson-distributed photon noise that is valid at our high photon count rates. We assume that each dataset corresponds to 5 hours of integration time, and thus generate 300 exposures per dataset. We generate a total of 1000 such datasets for analysis.

We also generate off-axis point-spread-functions (PSFs) that can be injected as astrophysical signal. The off-axis PSFs do not include WFE, since the simulations show that, at the WFE amplitudes of interest in our work, the planet coupling at separations of interest is not significantly impacted. In order to create data with an injected companion, the off-axis PSF at the desired separation is scaled appropriately based on the desired flux ratio, then added to each exposure of the simulated intensity of the on-axis source.

\begin{figure*}[t]
\begin{center}
	\includegraphics[scale = 0.55]{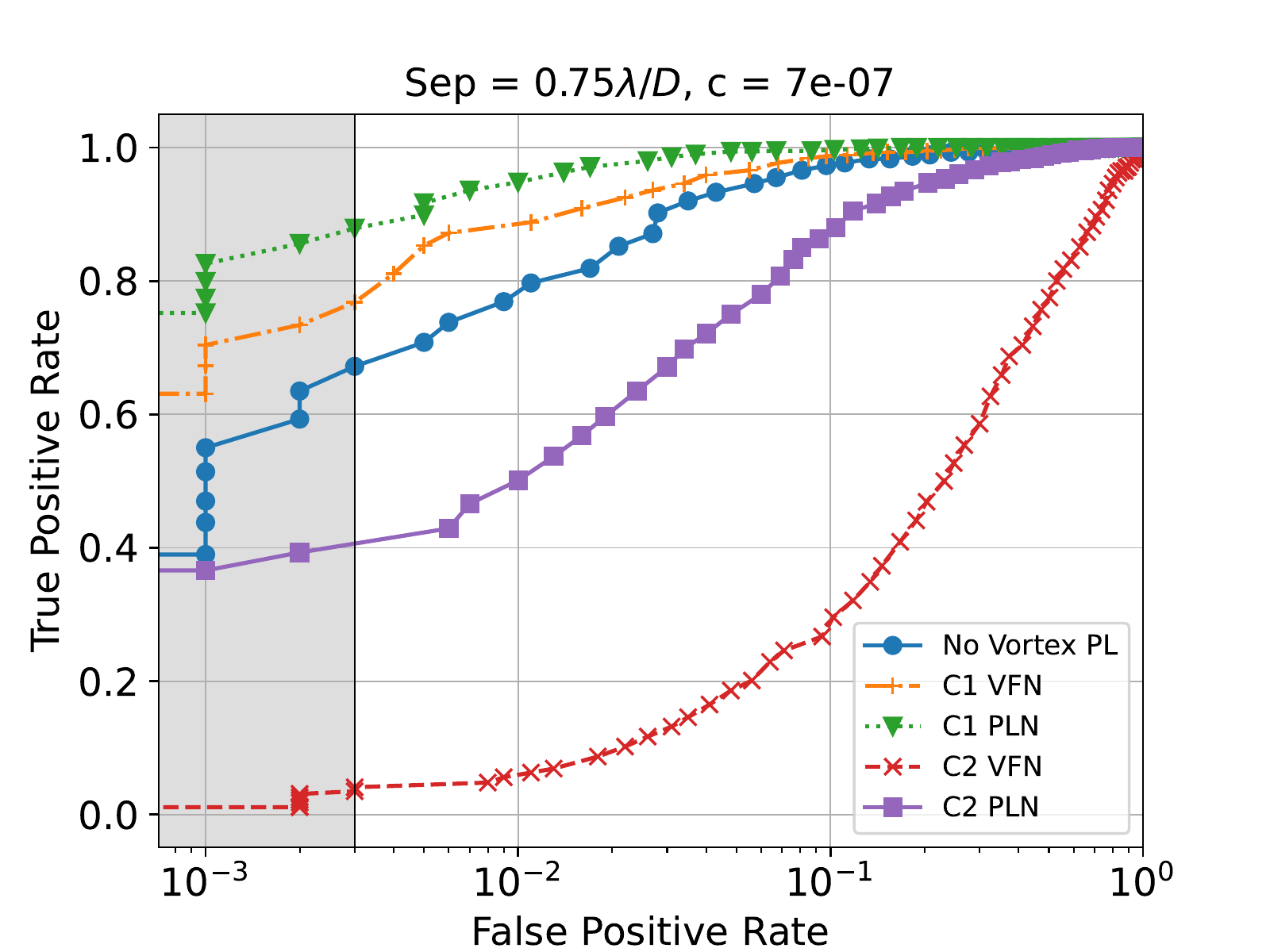}
	\includegraphics[scale = 0.55]{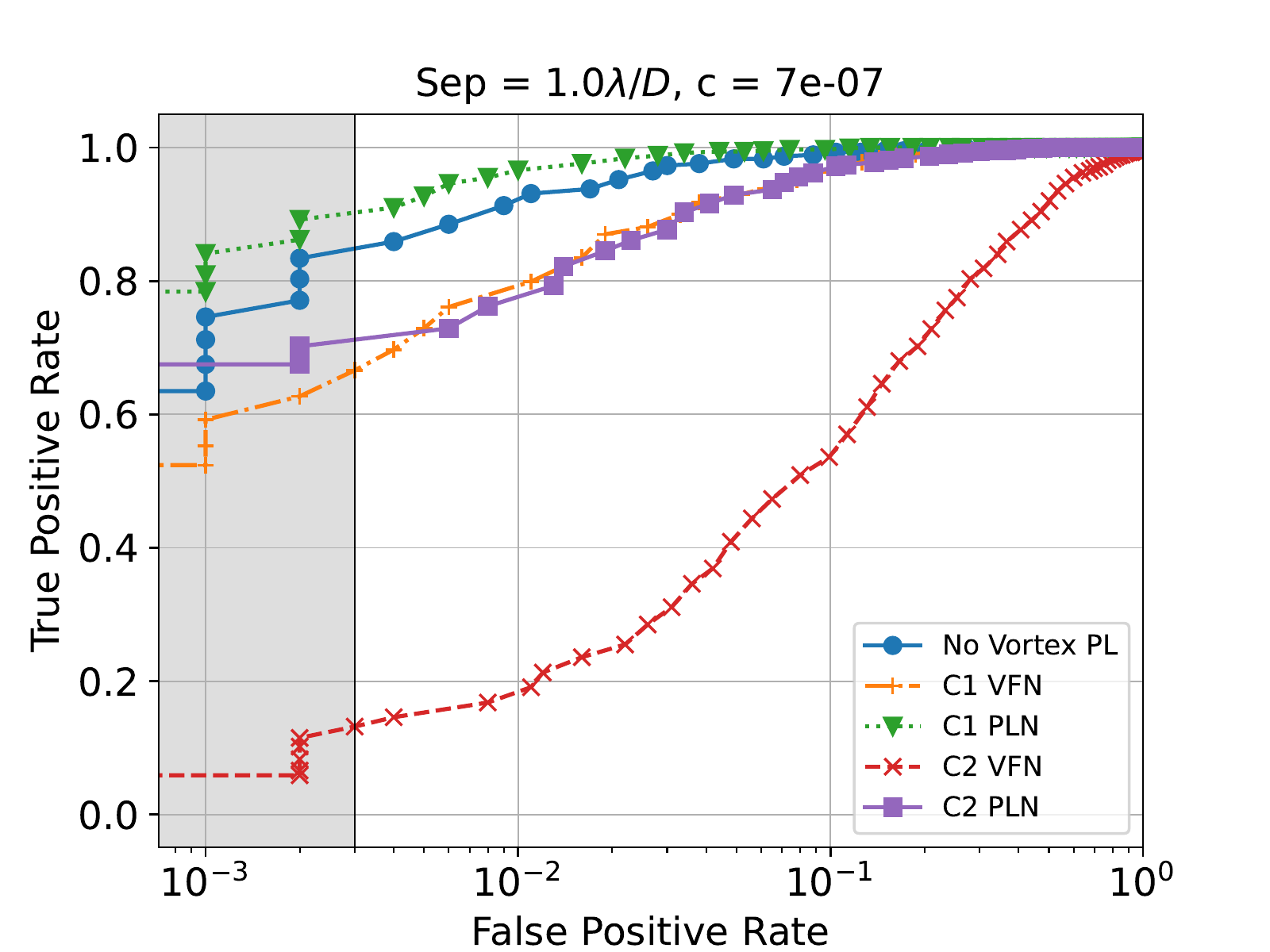}
	\includegraphics[scale = 0.55]{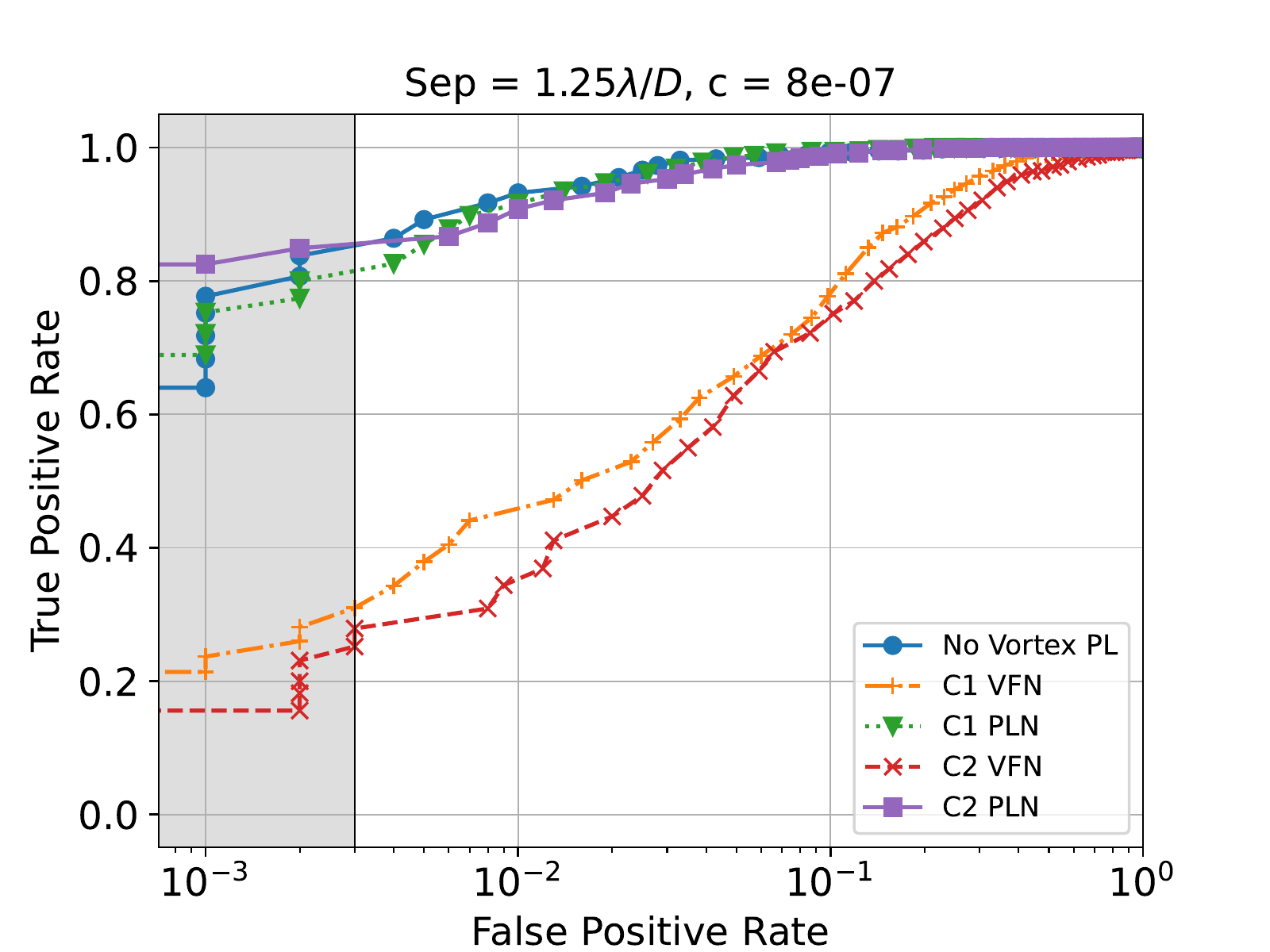}
	\includegraphics[scale = 0.55]{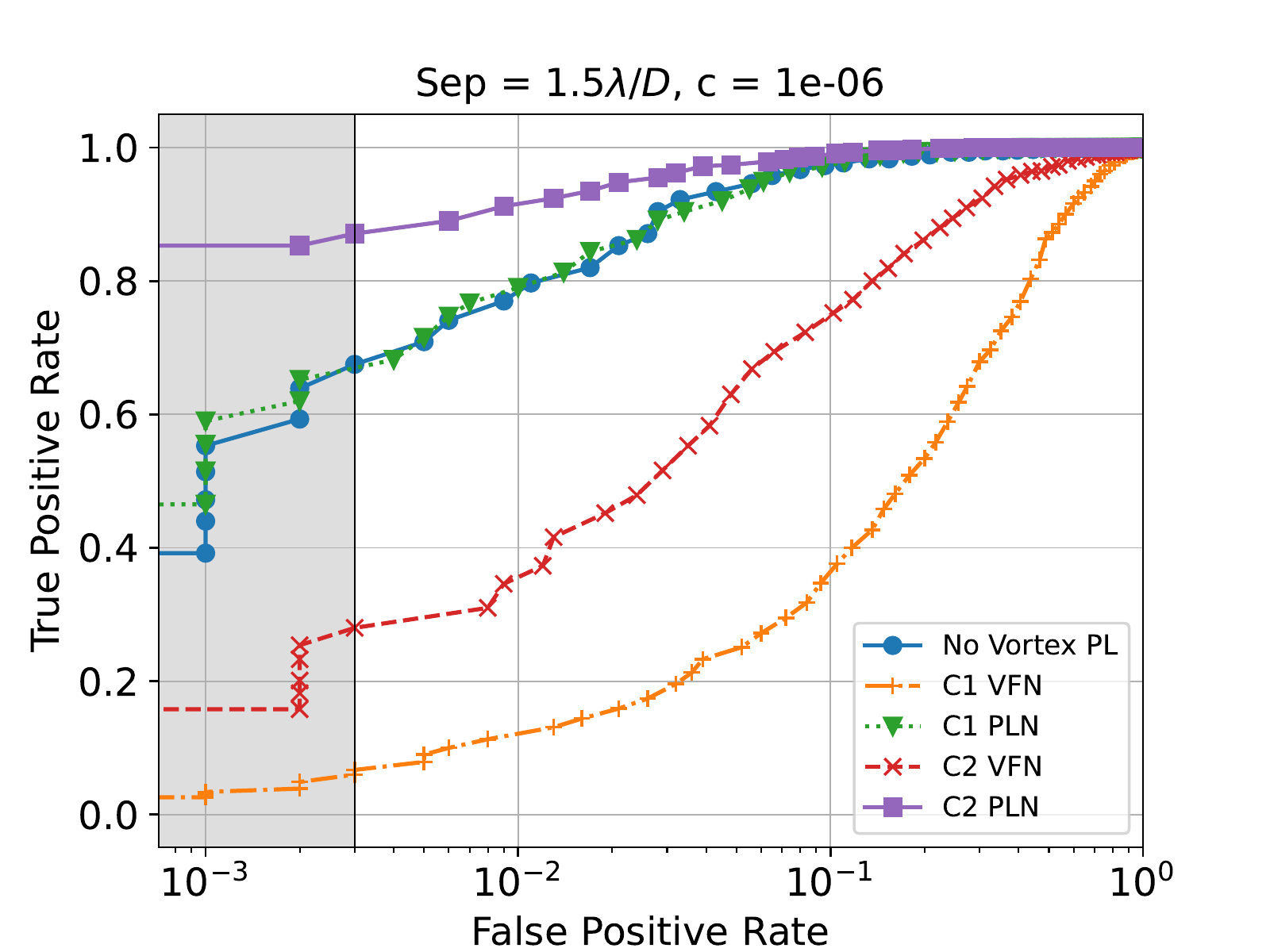}

	\caption{\label{fig:roc_curves} Example ROC curves at different separations in the presence of with photon noise, assuming wavefront error averages to a baseline null depth. For both vortex charges, the inclusion of other ports of the PLN provides detection gains relative to the VFN. The grey areas indicate false positive rates which are not well sampled as they involve fewer than 3 datasets with false detections.}
\end{center}
\end{figure*}

\subsection{Detection} \label{ssec:detection}

In this section, we characterize the detectability of planets, comparing the performance of the VFN and the PLN. For each dataset generated in Section \ref{sec:datagen}, we first take the mean of the 300 exposures and subtract off the nominal on-axis signal with no WFE. We then perform detection testing on the resulting data, using a total energy test statistic:

\begin{equation}
\label{eqn:energy_statistic}
\epsilon = \sum_i y_i^2,
\end{equation}

where $i$ is the port index of the PLN and $y_i$ the signal in the port. The test statistic $\epsilon$ is calculated from the data and compared to a threshold $\xi$, which is chosen to provide a desired false-alarm rate. A detection is claimed if $\epsilon \geq \xi$, and a lack of detection is claimed otherwise.

There are four possible outcomes when comparing the test statistic calculated from a dataset to the value of the test statistic set as the detection threshold. The first is a true positive, in that a real companion in the data is detected; the fraction of real companions detected is the true positive rate ($\mathrm{TPR}$). A second possible outcome is that a real companion is \textit{not} detected, occurring at a rate of $1-\mathrm{TPR}$. A third outcome is that there is no companion in the data, but the detection test incorrectly claims a detection. The rate at which this occurs is the false positive rate ($\mathrm{FPR}$). The fourth and last outcome is that there is no companion, and a detection is correctly not claimed, occurring at a rate of $1-\mathrm{FPR}$.

Choosing a threshold for the test statistic is a balancing act between the $\mathrm{TPR}$ and $\mathrm{FPR}$: as the threshold is decreased, detecting real companions becomes more likely, but false detections also become more likely. This dependency can be characterized by examining the possible values of the test statistic and calculating the $\mathrm{TPR}$ and $\mathrm{FPR}$ \textit{if} that value were the detection threshold. Plotting the $\mathrm{TPR}$ as a function of the $\mathrm{FPR}$ results in a receiver operating characteristic (ROC) curve, which characterizes the performance of a detection scheme and can be used in the determination of flux ratio detection limits.

Figure \ref{fig:roc_curves} shows ROC curves from the distribution of $\epsilon$ over the 1000 datasets. The VFN corresponds to the case where only the LP 01 port is used, while with the PLN, all four nulled ports are used. The simulations show that for both charges, the inclusion of the other nulled ports of the PLN provides detection gains relative to the VFN. For a given rate of false positives, the PLN can achieve a higher true positive rate than the VFN. At close in separations $\leq 1 \lambda/D$, the charge 1 PLN achieves the best performance. At separations greater than $\approx 1.25 \lambda/D$, the charge 2 PLN starts to perform better. Despite having higher throughput, the photonic lantern without a vortex does not outperform both the charge 1 and the charge 2 PLNs at any separation, emphasizing that the distribution of flux relative to the achievable null-depths matters more than sheer throughput. However, the no vortex PLN has the advantage of not requiring an additional optic in a pupil or focal plane, and can thus be realized with a simpler optical system. Additionally, the relative performance of the different configurations will ultimately depend on the distribution of WFE, as the ports in each configuration are sensitive to different subsets of modes.

\subsection{Model-Fitting}

Data from the VFN consists of only one measurement that contains no information on position angle and cannot discriminate between the effects of flux ratio and separation. Unlike the VFN, the spatial structures of the PLN modes allows for the retrieval of the planet's location, albeit with degeneracy in the position angle as a result of their symmetry.

To illustrate this capability, we attempt to fit models to one of the simulated datasets of the charge 2 VFN from Section \ref{sec:datagen}, where a planet with a flux ratio of $2\times 10^{-6} $ is injected at ($X=$ 1.25 $\lambda/D$, $Y=$ 0 $\lambda/D$). We believe that a configuration that slightly breaks the symmetry would be a better strategy for localization than any of the configurations presented here. Determining how to do this effectively would be part of future work. For this work, our primary aim was to show that this localization capability exists in this architecture, so we choose to focus on just one configuration.

First, we assume that the average null-depth can be estimated, such as by observing a reference star. This assumes telescope conditions are reasonably stable between observations of the reference and target stars, as the accuracy of the null-depth estimation will be impacted by quasi-static aberrations as well as differential alignment onto the vortex or lantern centers, which would lead to differences between the reference and target observations.

The estimated null-depth is subtracted from the average of the measurement frames. This step is necessary to debias the data, since if only the nominal on-axis signal (without any wavefront error) is subtracted, the WFE that sets the null-depth will contribute to the apparent flux of the planet. We then fit a model to the data through Chi-squared ($\chi^2$) minimization, using only data from the LP 01 port for the VFN, and data from all six ports for the PLN.

The three model parameters for a planet are its location coordinates $(X,Y)$ and its flux ratio (FR). We first generate a grid of parameter values, choosing $X$ to span from 0 $\lambda/D$ to 3 $\lambda/D$ and $Y$ to span from -3 $\lambda/D$ to 3 $\lambda/D$. This spans the spatial half-plane, which is enough for our purposes, as the symmetry of the modes means the position angle can at best be localized with a 180 degeneracy. The flux ratios are chosen to range logarithmically from $10^{-7}$ to $10^{-5}$.

A planet corresponding to each set of parameters from the grid is simulated with the instrument model. The $\chi^2$ of the difference between the model and the data is calculated using $\chi^2 = \sum_i (y_i - x_i)^2/\sigma_i^2$, where $y_i$ is the measured data in port $i$, $x_i$ is the model, and $\sigma_i$ is the standard deviation of the noise across the 300 frames. The probability distribution is then calculated by taking $\mathrm{P}(X,Y,\mathrm{FR}) \propto \exp{-\chi^2/2}$, and normalizing such that the total probability over the entire explored parameter space is 1.

\begin{figure*}[t]
\begin{center}
	\fbox{\includegraphics[scale = 0.35]{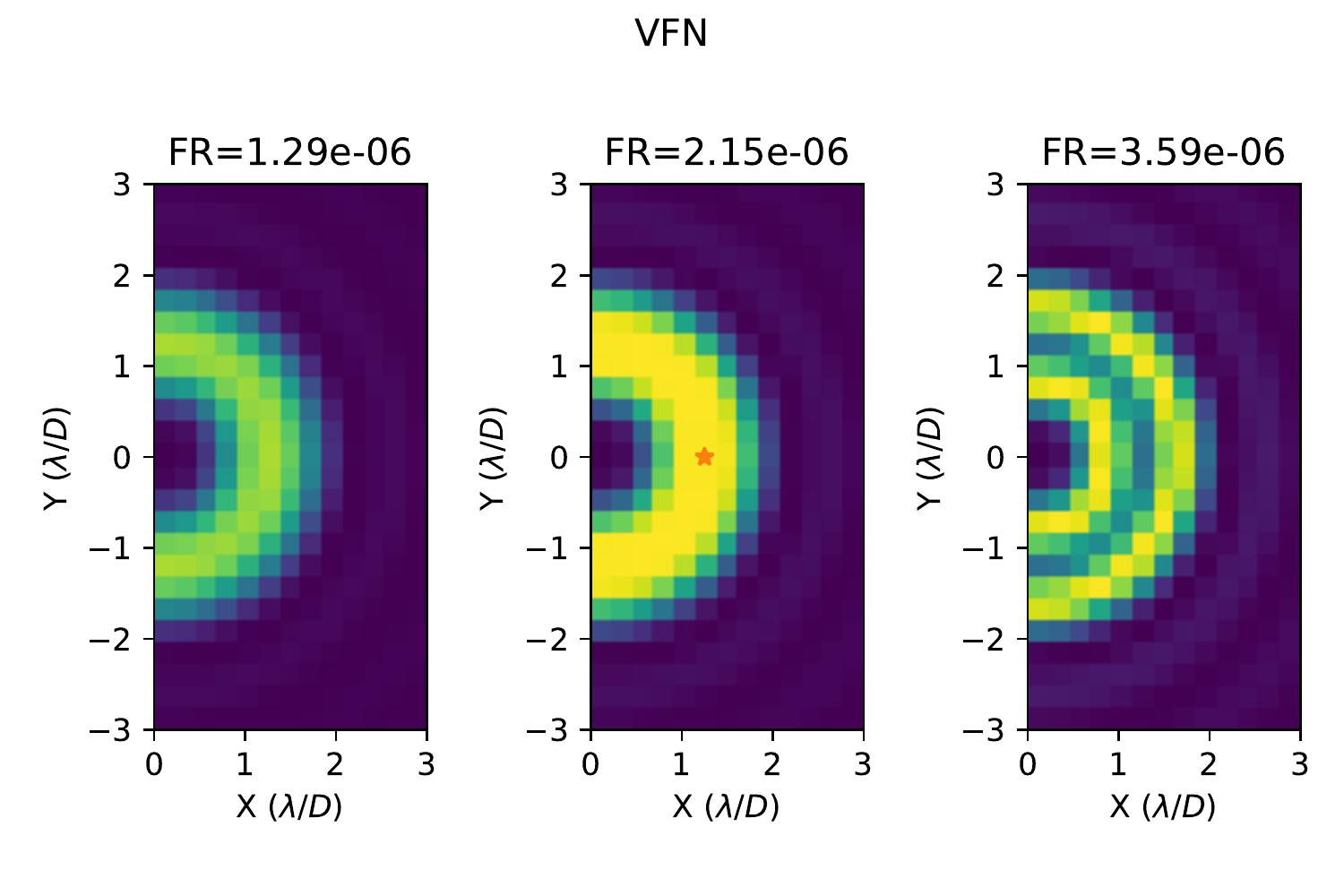}}
	\fbox{\includegraphics[scale = 0.35]{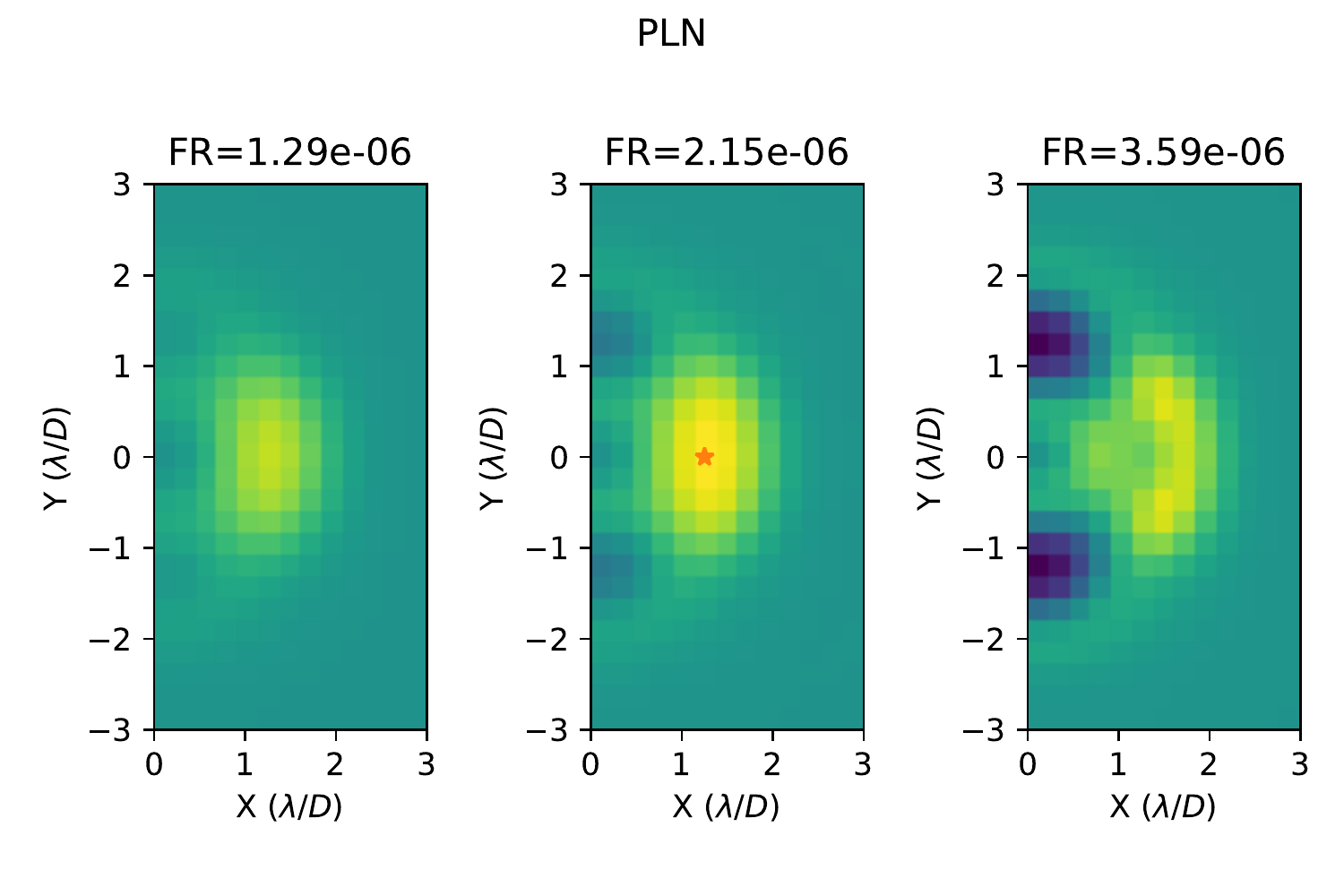}}
	\includegraphics[width=0.34\textwidth,height=120pt]{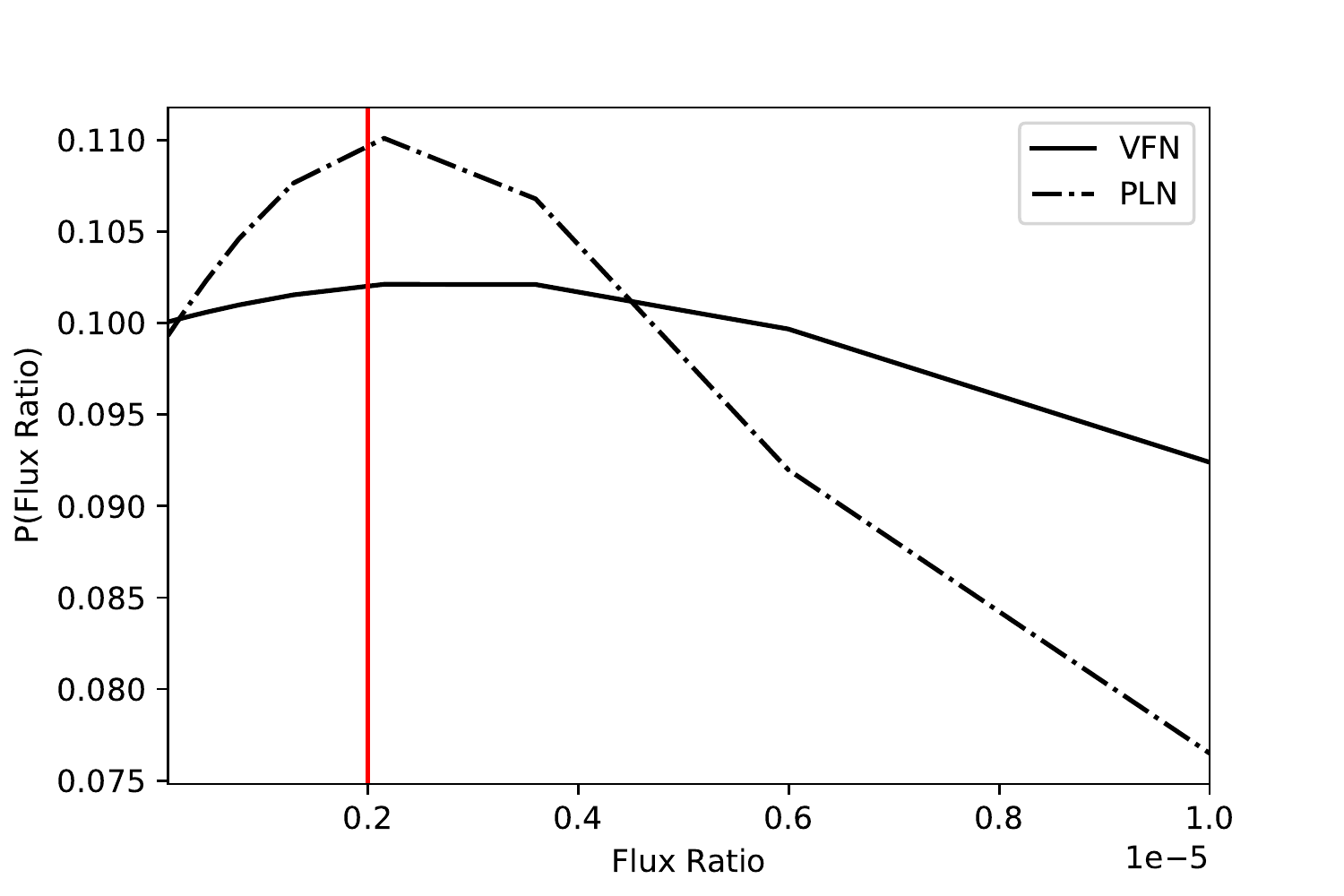}
	\caption{\label{fig:probmaps} Left: Select spatial probability distribution cross-sections, using a charge 2 VFN. The three panels are plotted on the same color scale. Middle: Select spatial probability distribution cross-sections, using a charge 2 PLN. The three panels are plotted on the same color scale. The parameters closest to that of the injected planet are marked with orange stars. Right: Probability distributions of the flux ratio, marginalized over the spatial dimensions. The flux ratio of the injected planet is marked by the red line. The model-fitting shows that the PLN can provide better constraints on planet model parameters compared to the VFN.}
\end{center}
\end{figure*}

Figure \ref{fig:probmaps} depicts the three spatial cross-sections of the resulting probability distributions for the charge 2 VFN and PLN, corresponding to the flux ratio values from the grid closest to the injected value of $2\times 10^{-6}$. The parameter set in the grid closest to that of the injected planet is marked with an orange star. Also shown is the probability distribution of the flux ratio, marginalized over the spatial dimensions. As expected, it is largely unconstrained by the VFN, which cannot distinguish between the competing effects of flux ratio and separation. However, with the spatial information provided by the PLN, the retrieved probability distribution of the flux ratio peaks at the correct value of $2\times10^{-6}$. Given the best fit flux ratio using PLN, fitting a gaussian curve to the y-axis cross-section of the spatial probability distribution reveals that the position angle can be localized to $\sim 1~\lambda/D$ with the PLN, while it is completely unconstrained by the VFN. These simulation results show that compared to the VFN, the PLN can provide better constraints on the planet's location and flux ratio. 

\begin{figure*}[t]
\begin{center}
	\includegraphics[scale = 0.5]{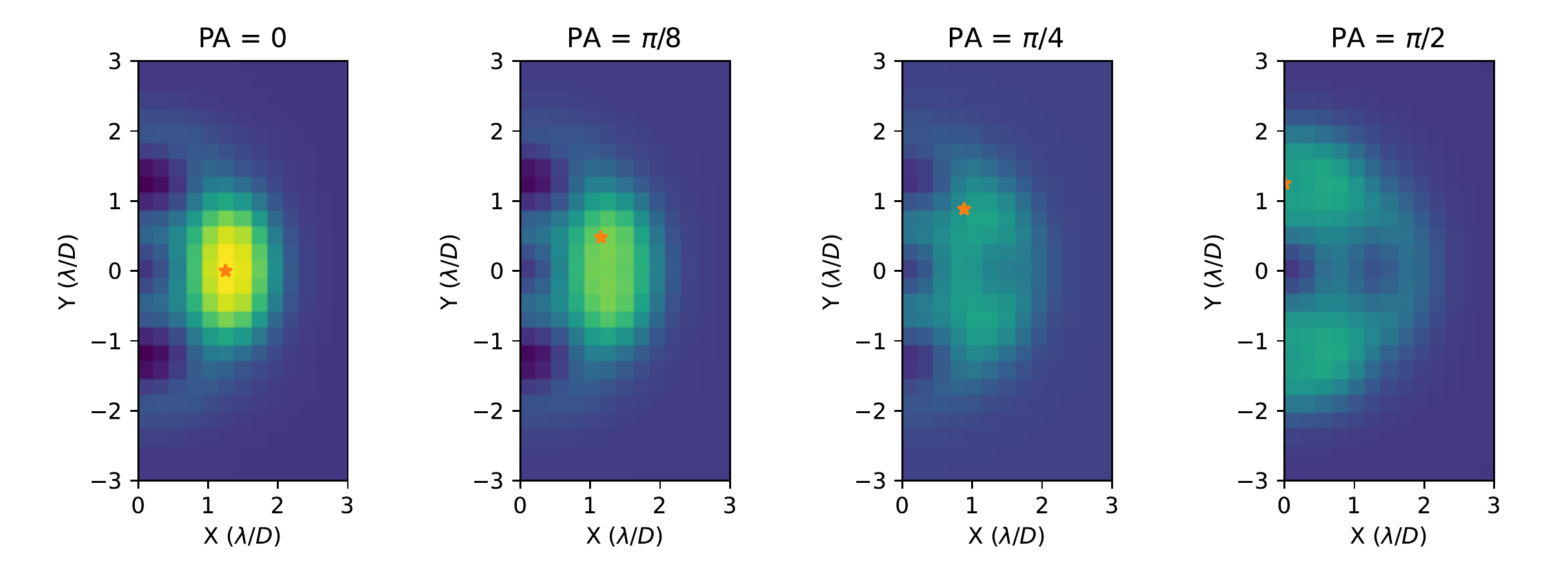}
	\caption{\label{fig:pa_scan} Spatial probability distributions given the correctly identified flux ratio of $2.15\times10^{-6}$ (the panels are plotted on the same color scale). Planets at a separation of 1.25 $\lambda/D$ are injected at a variety of injected position angles (marked by the orange stars). At position angles other than 0 and $\pi/2$, additional solutions exist beyond the two guaranteed by the instrumental symmetry. However, an observing strategy that involves taking data with multiple rotations of the instrument relative to the sky will reduce the number of best fit position angle solutions to the fundamental two.
	}
\end{center}
\end{figure*}

The response of the PLN to off-axis signal is not rotationally symmetric. We thus explore injecting and recovering a planet signal at varying position angles. Figure \ref{fig:pa_scan} shows that, given the correct flux ratio, the localization response varies as a function of position angle. At position angles other than 0 and $\pi/2$, additional solutions exist beyond the two guaranteed by the instrumental symmetry. However, an observing strategy that involves taking data with multiple rotations of the instrument relative to the sky will reduce the number of best fit position angle solutions to the fundamental two. Finding the most efficient observational strategy to best constrain the position angle given an unknown random initial orientation, and exploring the possibility of introducing slight asymmetries to break this degeneracy, are topics left for future work.

\newpage
\section{Conclusions}

This work presents a proof-of-concept study of the Photonic Lantern Vortex Fiber Nuller. The advantage the MSPL offers over the SMF is two-fold. First, a photonic lantern, regardless of modal selectivity, accepts more input modes than the SMF, increasing the overall amount of light that can couple in. This improves the overall field of view and total planet coupling provided by the VFN. Second, the symmetries resulting from modal selectivity interact with the vortex field to create not just on-axis nulls, but also ports insensitive to low-order aberrations that do not meet a specific azimuthal order condition. Together, these properties of the PLN result in an instrument that rejects starlight while maintaining a substantial amount of planet light in the regions of interest. Additionally, while the PLN is meant for integration with spectrographs, motivated by the science that can be done in the spectral domain, the ports with different modal structures captures some spatial information, enabling planet localization that is not possible with the VFN. However, the instrumental symmetries that provide starlight and wavefront error rejection currently also cause degeneracies in the spatial information captured. Future work will explore whether introducing slight asymmetries into the instrument can lift the spatial degeneracies with minimal impact to the achievable null depth.

This work simulates the PLN's ideal behavior at a single wavelength. However, the modes of a realistic mode-selective photonic lantern will deviate from the ideal LP modes. Furthermore, its modes will actually vary with wavelength. Finite-difference beam propagation simulations are needed to simulate the behavior of a realistic photonic lantern design across different wavelengths, since its modes will no longer correspond to perfect LP modes, and there will be modal cross-coupling due to imperfections in the design as well as the fabrication process. Additional performance simulations will be conducted to characterize the impact of this non-ideal, wavelength dependent behavior on science results. This work includes simulating the PLN with synthetic planetary spectra and investigating methods to analyze the data, building upon current practices in exoplanet spectral analysis \citep{Wang_2021}. We will identify best practices to account for the wavelength dependent mode-structure and throughput and the optimal method for combining data from the different ports, including the possibility of obtaining concurrent stellar spectra in the non-nulled ports to be used for calibration and analysis. We will investigate if multiple sets of spectroscopic data can be used to cross-calibrate systematic errors. The single-mode outputs are ideal for downstream spectroscopy using photonic spectrographs \citep{gatkine2019astrophotonic}. 
We will thus investigate strategies for optimal integration of PLN with an on-chip photonic spectrograph on each of the single-mode outputs (nulled or otherwise) to measure the spectra of the planet/companion and star, as well for cross-calibration.

Future work also includes verifying the behavior of a PLN in the lab — both the characterization of the photonic lantern device itself, and after integration with a vortex. We intend to characterize the PLN with different levels of wavefront error, as well as investigate the possibility of performing wavefront control to achieve better nulls, potentially compensating for defects such as residual optical surface error or even non-ideal photonic lantern modes. If the laboratory characterization validates the performance of the PLN, an on-sky demonstration will be attempted.

This work on the PLN also naturally ties in to several related topics, such as the development of wavefront sensing algorithms through photonic lanterns \citep{Lin_PLWFS1,Norris_2020}, or the leveraging of the photonic lantern design paradigm to push towards the theoretical limits of optical signal separation.

\acknowledgments


We thank the anonymous reviewer for their careful consideration and feedback. This work is supported in part by the National Science Foundation Graduate Research Fellowship under Grant No. 1122374.  Additional effort has been supported by the National Science Foundation under Grant No. 2109231. This research was carried out in part at the California Institute of Technology and the Jet Propulsion Laboratory under a contract with the National Aeronautics and Space Administration (NASA).

%

\vspace{5mm}

\software{This research made use of Astropy \citep{astropy:2013, astropy:2018}; NumPy \citep{oliphant2006guide_numpy}; SciPy \citep{2020SciPy-NMeth}; and Matplotlib \citep{Hunter:2007_matplotlib}.}




\pagebreak
\appendix

\section{Timeseries of Instantaneous Coupling with KPIC Residuals} \label{app:timeseries}
\begin{figure}[ht]
\begin{center}
    No Vortex\\
	\includegraphics[scale = 0.35]{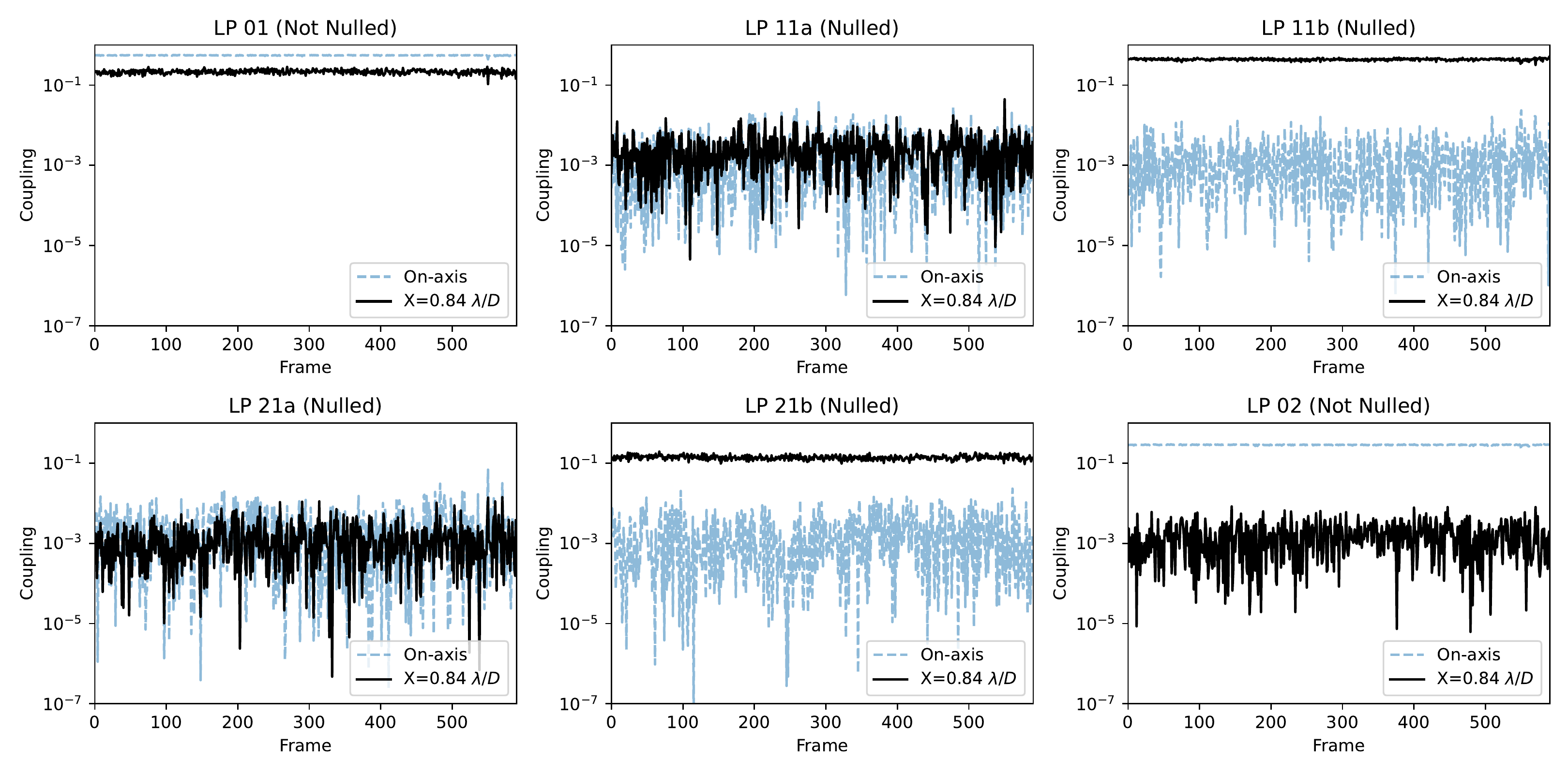}\\
    Charge 1\\
	\includegraphics[scale = 0.35]{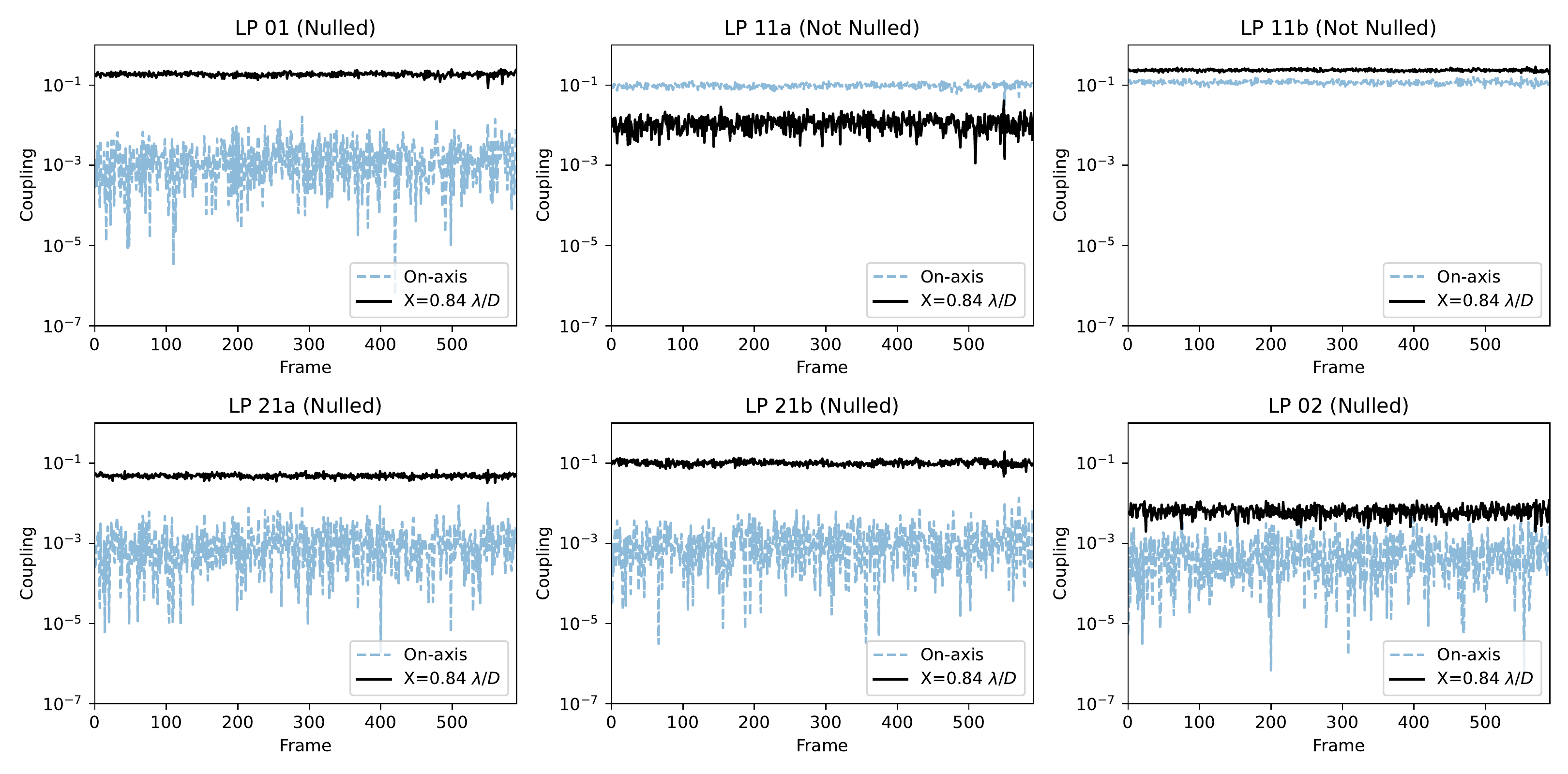}\\
	Charge 2\\
	\includegraphics[scale = 0.35]{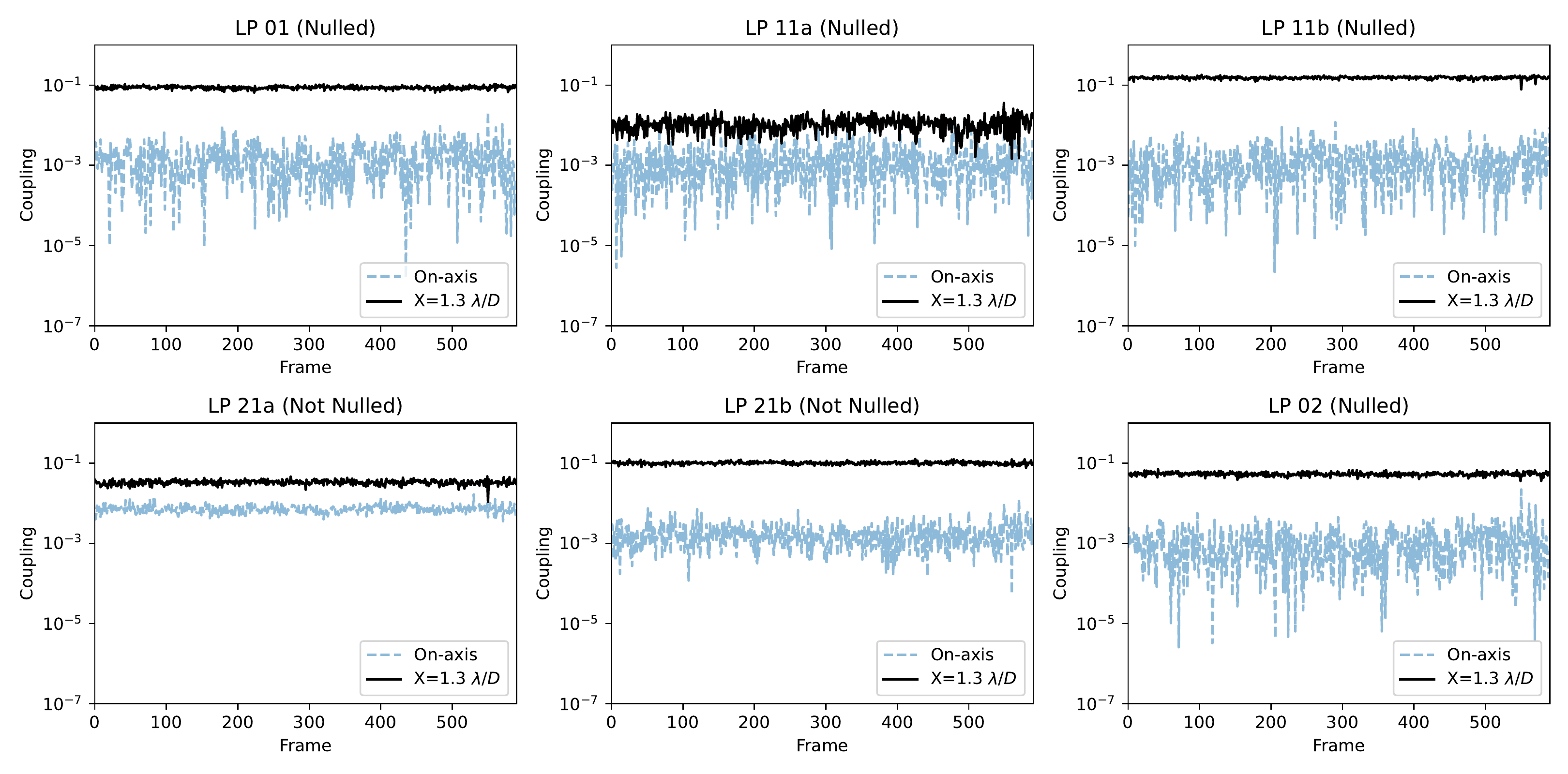}
	\caption{\label{fig:kpic_coupling_timeseries} Coupling calculated over 590 frames of WFE residuals from the KPIC PyWFS, with no vortex (top), a charge 1 vortex (middle), and a charge 2 vortex (bottom). Off-axis planet coupling (where it is expected based on coupling maps) remains higher than the stellar coupling in the presence of these WFE realizations.}
\end{center}
\end{figure}


\bibliography{biblio}{}

\begin{thebibliography}{}
\expandafter\ifx\csname natexlab\endcsname\relax\def\natexlab#1{#1}\fi
\providecommand{\url}[1]{\href{#1}{#1}}
\providecommand{\dodoi}[1]{doi:~\href{http://doi.org/#1}{\nolinkurl{#1}}}
\providecommand{\doeprint}[1]{\href{http://ascl.net/#1}{\nolinkurl{http://ascl.net/#1}}}
\providecommand{\doarXiv}[1]{\href{https://arxiv.org/abs/#1}{\nolinkurl{https://arxiv.org/abs/#1}}}

\bibitem[{{Astropy Collaboration} {et~al.}(2013){Astropy Collaboration},
  {Robitaille}, {Tollerud}, {Greenfield}, {Droettboom}, {Bray}, {Aldcroft},
  {Davis}, {Ginsburg}, {Price-Whelan}, {Kerzendorf}, {Conley}, {Crighton},
  {Barbary}, {Muna}, {Ferguson}, {Grollier}, {Parikh}, {Nair}, {Unther},
  {Deil}, {Woillez}, {Conseil}, {Kramer}, {Turner}, {Singer}, {Fox}, {Weaver},
  {Zabalza}, {Edwards}, {Azalee Bostroem}, {Burke}, {Casey}, {Crawford},
  {Dencheva}, {Ely}, {Jenness}, {Labrie}, {Lim}, {Pierfederici}, {Pontzen},
  {Ptak}, {Refsdal}, {Servillat}, \& {Streicher}}]{astropy:2013}
{Astropy Collaboration}, {Robitaille}, T.~P., {Tollerud}, E.~J., {et~al.} 2013,
  Astronomy and Astrophysics, 558, A33, \dodoi{10.1051/0004-6361/201322068}

\bibitem[{{Bracewell}(1978)}]{bracewell_1978}
{Bracewell}, R.~N. 1978, \nat, 274, 780, \dodoi{10.1038/274780a0}

\bibitem[{Corrigan {et~al.}(2018)Corrigan, Morris, Harris, \&
  Anagnos}]{corrigan2018}
Corrigan, M.~K., Morris, T.~J., Harris, R.~J., \& Anagnos, T. 2018, in Adaptive
  Optics Systems VI, ed. L.~M. Close, L.~Schreiber, \& D.~Schmidt, Vol. 10703,
  International Society for Optics and Photonics (SPIE), 1313 -- 1320,
  \dodoi{10.1117/12.2311336}

\bibitem[{{Delorme} {et~al.}(2021){Delorme}, {Jovanovic}, {Echeverri}, {Mawet},
  {Kent Wallace}, {Bartos}, {Cetre}, {Wizinowich}, {Ragland}, {Lilley},
  {Wetherell}, {Doppmann}, {Wang}, {Morris}, {Ruffio}, {Martin}, {Fitzgerald},
  {Ruane}, {Schofield}, {Suominen}, {Calvin}, {Wang}, {Magnone}, {Johnson},
  {Sohn}, {L{\'o}pez}, {Bond}, {Pezzato}, {Sayson}, {Chun}, \&
  {Skemer}}]{delorme_2021}
{Delorme}, J.-R., {Jovanovic}, N., {Echeverri}, D., {et~al.} 2021, Journal of
  Astronomical Telescopes, Instruments, and Systems, 7, 035006,
  \dodoi{10.1117/1.JATIS.7.3.035006}

\bibitem[{{Echeverri} {et~al.}(2019){Echeverri}, {Ruane}, {Jovanovic}, {Mawet},
  \& {Levraud}}]{Echeverri_VFN}
{Echeverri}, D., {Ruane}, G., {Jovanovic}, N., {Mawet}, D., \& {Levraud}, N.
  2019, Optics Letters, 44, 2204, \dodoi{10.1364/OL.44.002204}

\bibitem[{Echeverri {et~al.}(2019)Echeverri, Ruane, Jovanovic, Hayama, Delorme,
  Pezzato, Bond, Wang, Mawet, Wallace, \& Serabyn}]{echeverri_spie_2019}
Echeverri, D., Ruane, G., Jovanovic, N., {et~al.} 2019, in Techniques and
  Instrumentation for Detection of Exoplanets IX, ed. S.~B. Shaklan, Vol.
  11117, International Society for Optics and Photonics (SPIE), 264 -- 274,
  \dodoi{10.1117/12.2528529}

\bibitem[{Gatkine {et~al.}(2019)Gatkine, Veilleux, \&
  Dagenais}]{gatkine2019astrophotonic}
Gatkine, P., Veilleux, S., \& Dagenais, M. 2019, Applied Sciences, 9, 290

\bibitem[{Hunter(2007)}]{Hunter:2007_matplotlib}
Hunter, J.~D. 2007, Computing in Science \& Engineering, 9, 90,
  \dodoi{10.1109/MCSE.2007.55}

\bibitem[{{Jovanovic} {et~al.}(2016){Jovanovic}, {Schwab}, {Cvetojevic},
  {Guyon}, \& {Martinache}}]{jovanovic-ESS-2016}
{Jovanovic}, N., {Schwab}, C., {Cvetojevic}, N., {Guyon}, O., \& {Martinache},
  F. 2016, \pasp, 128, 121001, \dodoi{10.1088/1538-3873/128/970/121001}

\bibitem[{{Leon-Saval} {et~al.}(2013){Leon-Saval}, {Argyros}, \&
  {Bland-Hawthorn}}]{LeonSaval_PL_2013}
{Leon-Saval}, S.~G., {Argyros}, A., \& {Bland-Hawthorn}, J. 2013,
  Nanophotonics, 2, 429, \dodoi{10.1515/nanoph-2013-0035}

\bibitem[{Leon-Saval {et~al.}(2014)Leon-Saval, Fontaine, Salazar-Gil, Ercan,
  Ryf, \& Bland-Hawthorn}]{LeonSaval_MSPL}
Leon-Saval, S.~G., Fontaine, N.~K., Salazar-Gil, J.~R., {et~al.} 2014, Opt.
  Express, 22, 1036, \dodoi{10.1364/OE.22.001036}

\bibitem[{Lin {et~al.}(2022)Lin, Fitzgerald, Xin, Guyon, Leon-Saval, Norris, \&
  Jovanovic}]{Lin_PLWFS1}
Lin, J., Fitzgerald, M., Xin, Y., {et~al.} 2022, {in prep.}

\bibitem[{{Lin} {et~al.}(2021){Lin}, {Jovanovic}, \& {Fitzgerald}}]{lin_2021}
{Lin}, J., {Jovanovic}, N., \& {Fitzgerald}, M.~P. 2021, Journal of the Optical
  Society of America B Optical Physics, 38, A51, \dodoi{10.1364/JOSAB.423664}

\bibitem[{{{National Research Council}}(2021)}]{NRC_2020Decadal}
{{National Research Council}}. 2021, Pathways to Discovery in Astronomy and
  Astrophysics for the 2020s (Washington, DC: The National Academies Press)

\bibitem[{{Norris} {et~al.}(2020){Norris}, {Wei}, {Betters}, {Wong}, \&
  {Leon-Saval}}]{Norris_2020}
{Norris}, B. R.~M., {Wei}, J., {Betters}, C.~H., {Wong}, A., \& {Leon-Saval},
  S.~G. 2020, Nature Communications, 11, 5335,
  \dodoi{10.1038/s41467-020-19117-w}

\bibitem[{Oliphant(2006)}]{oliphant2006guide_numpy}
Oliphant, T.~E. 2006, A guide to NumPy, Vol.~1 (Trelgol Publishing USA)

\bibitem[{Paschotta(2022{\natexlab{a}})}]{lp_mode_def}
Paschotta, R. 2022{\natexlab{a}}, LP modes,  RP Photonics.
\newblock \url{https://www.rp-photonics.com/lp_modes.html}

\bibitem[{Paschotta(2022{\natexlab{b}})}]{vnum_def}
---. 2022{\natexlab{b}}, V number,  RP Photonics.
\newblock \url{https://www.rp-photonics.com/v_number.html}

\bibitem[{Por {et~al.}(2018)Por, Haffert, Radhakrishnan, Doelman, Van~Kooten,
  \& Bos}]{por2018hcipy}
Por, E.~H., Haffert, S.~Y., Radhakrishnan, V.~M., {et~al.} 2018, in Proc.
  {{SPIE}}, Vol. 10703, Adaptive Optics Systems VI, \dodoi{10.1117/12.2314407}

\bibitem[{{Price-Whelan} {et~al.}(2018){Price-Whelan}, {Sip{\H{o}}cz},
  {G{\"u}nther}, {Lim}, {Crawford}, {Conseil}, {Shupe}, {Craig}, {Dencheva},
  {Ginsburg}, {VanderPlas}, {Bradley}, {P{\'e}rez-Su{\'a}rez}, {de Val-Borro},
  {Paper Contributors}, {Aldcroft}, {Cruz}, {Robitaille}, {Tollerud},
  {Coordination Committee}, {Ardelean}, {Babej}, {Bach}, {Bachetti}, {Bakanov},
  {Bamford}, {Barentsen}, {Barmby}, {Baumbach}, {Berry}, {Biscani}, {Boquien},
  {Bostroem}, {Bouma}, {Brammer}, {Bray}, {Breytenbach}, {Buddelmeijer},
  {Burke}, {Calderone}, {Cano Rodr{\'\i}guez}, {Cara}, {Cardoso}, {Cheedella},
  {Copin}, {Corrales}, {Crichton}, {D{\textquoteright}Avella}, {Deil},
  {Depagne}, {Dietrich}, {Donath}, {Droettboom}, {Earl}, {Erben}, {Fabbro},
  {Ferreira}, {Finethy}, {Fox}, {Garrison}, {Gibbons}, {Goldstein}, {Gommers},
  {Greco}, {Greenfield}, {Groener}, {Grollier}, {Hagen}, {Hirst}, {Homeier},
  {Horton}, {Hosseinzadeh}, {Hu}, {Hunkeler}, {Ivezi{\'c}}, {Jain}, {Jenness},
  {Kanarek}, {Kendrew}, {Kern}, {Kerzendorf}, {Khvalko}, {King}, {Kirkby},
  {Kulkarni}, {Kumar}, {Lee}, {Lenz}, {Littlefair}, {Ma}, {Macleod},
  {Mastropietro}, {McCully}, {Montagnac}, {Morris}, {Mueller}, {Mumford},
  {Muna}, {Murphy}, {Nelson}, {Nguyen}, {Ninan}, {N{\"o}the}, {Ogaz}, {Oh},
  {Parejko}, {Parley}, {Pascual}, {Patil}, {Patil}, {Plunkett}, {Prochaska},
  {Rastogi}, {Reddy Janga}, {Sabater}, {Sakurikar}, {Seifert}, {Sherbert},
  {Sherwood-Taylor}, {Shih}, {Sick}, {Silbiger}, {Singanamalla}, {Singer},
  {Sladen}, {Sooley}, {Sornarajah}, {Streicher}, {Teuben}, {Thomas},
  {Tremblay}, {Turner}, {Terr{\'o}n}, {van Kerkwijk}, {de la Vega}, {Watkins},
  {Weaver}, {Whitmore}, {Woillez}, {Zabalza}, \& {Contributors}}]{astropy:2018}
{Price-Whelan}, A.~M., {Sip{\H{o}}cz}, B.~M., {G{\"u}nther}, H.~M., {et~al.}
  2018, Astronomical Journal, 156, 123, \dodoi{10.3847/1538-3881/aabc4f}

\bibitem[{Ruane {et~al.}(2019)Ruane, Echeverri, Jovanovic, Mawet, Serabyn,
  Wallace, Wang, \& Batalha}]{ruane_2019_spie}
Ruane, G., Echeverri, D., Jovanovic, N., {et~al.} 2019, in Techniques and
  Instrumentation for Detection of Exoplanets IX, ed. S.~B. Shaklan, Vol.
  11117, International Society for Optics and Photonics (SPIE), 366 -- 381,
  \dodoi{10.1117/12.2528555}

\bibitem[{Ruane {et~al.}(2018)Ruane, Wang, Mawet, {et~al.}}]{Ruane2018_VFN}
Ruane, G., Wang, J., Mawet, D., {et~al.} 2018, Astrophys. J.

\bibitem[{{Serabyn} {et~al.}(2019){Serabyn}, {Mennesson}, {Martin}, {Liewer},
  \& {K{\"u}hn}}]{pfn}
{Serabyn}, E., {Mennesson}, B., {Martin}, S., {Liewer}, K., \& {K{\"u}hn}, J.
  2019, \mnras, 489, 1291, \dodoi{10.1093/mnras/stz2163}

\bibitem[{{Stark} {et~al.}(2015){Stark}, {Roberge}, {Mandell}, {Clampin},
  {Domagal-Goldman}, {McElwain}, \& {Stapelfeldt}}]{Stark_2015}
{Stark}, C.~C., {Roberge}, A., {Mandell}, A., {et~al.} 2015, \apj, 808, 149,
  \dodoi{10.1088/0004-637X/808/2/149}

\bibitem[{Tuthill(2022)}]{Tuthill2022-NIH}
Tuthill, P. 2022, in Adaptive Optics Systems VIII, ed. L.~Schreiber,
  D.~Schmidt, \& E.~Vernet, Vol. 12185, International Society for Optics and
  Photonics (SPIE), 121858P, \dodoi{10.1117/12.2627887}

\bibitem[{Tuthill {et~al.}(2000)Tuthill, Monnier, \& Danchi}]{nrm_tuthill}
Tuthill, P.~G., Monnier, J.~D., \& Danchi, W.~C. 2000, in Interferometry in
  Optical Astronomy, ed. P.~J. Lena \& A.~Quirrenbach, Vol. 4006, International
  Society for Optics and Photonics (SPIE), 491 -- 498,
  \dodoi{10.1117/12.390244}

\bibitem[{Vel{\'a}zquez-Ben{\'i}tez {et~al.}(2018)Vel{\'a}zquez-Ben{\'i}tez,
  Antonio-L{\'o}pez, Alvarado-Zacar{\'i}as, Fontaine, Ryf, Chen,
  Hern{\'a}ndez-Cordero, Sillard, Okonkwo, Leon-Saval, \&
  Amezcua-Correa}]{Velazquez-Benitez2018}
Vel{\'a}zquez-Ben{\'i}tez, A.~M., Antonio-L{\'o}pez, J.~E.,
  Alvarado-Zacar{\'i}as, J.~C., {et~al.} 2018, Scientific Reports, 8, 8897,
  \dodoi{10.1038/s41598-018-27072-2}

\bibitem[{{Virtanen} {et~al.}(2020){Virtanen}, {Gommers}, {Oliphant},
  {Haberland}, {Reddy}, {Cournapeau}, {Burovski}, {Peterson}, {Weckesser},
  {Bright}, {van der Walt}, {Brett}, {Wilson}, {Jarrod Millman}, {Mayorov},
  {Nelson}, {Jones}, {Kern}, {Larson}, {Carey}, {Polat}, {Feng}, {Moore}, {Vand
  erPlas}, {Laxalde}, {Perktold}, {Cimrman}, {Henriksen}, {Quintero}, {Harris},
  {Archibald}, {Ribeiro}, {Pedregosa}, {van Mulbregt}, \&
  {Contributors}}]{2020SciPy-NMeth}
{Virtanen}, P., {Gommers}, R., {Oliphant}, T.~E., {et~al.} 2020, Nature
  Methods, 17, 261, \dodoi{https://doi.org/10.1038/s41592-019-0686-2}

\bibitem[{{Wang} {et~al.}(2017){Wang}, {Mawet}, {Ruane}, {Hu}, \&
  {Benneke}}]{wang_ji_2017}
{Wang}, J., {Mawet}, D., {Ruane}, G., {Hu}, R., \& {Benneke}, B. 2017, \aj,
  153, 183, \dodoi{10.3847/1538-3881/aa6474}

\bibitem[{{Wang} {et~al.}(2021){Wang}, {Ruffio}, {Morris}, {Delorme},
  {Jovanovic}, {Pezzato}, {Echeverri}, {Finnerty}, {Hood}, {Zanazzi}, {Bryan},
  {Bond}, {Cetre}, {Martin}, {Mawet}, {Skemer}, {Baker}, {Xuan}, {Wallace},
  {Wang}, {Bartos}, {Blake}, {Boden}, {Buzard}, {Calvin}, {Chun}, {Doppmann},
  {Dupuy}, {Duch{\^e}ne}, {Feng}, {Fitzgerald}, {Fortney}, {Freedman},
  {Knutson}, {Konopacky}, {Lilley}, {Liu}, {Lopez}, {Lupu}, {Marley},
  {Meshkat}, {Miles}, {Millar-Blanchaer}, {Ragland}, {Roy}, {Ruane}, {Sappey},
  {Schofield}, {Weiss}, {Wetherell}, {Wizinowich}, \& {Ygouf}}]{Wang_2021}
{Wang}, J.~J., {Ruffio}, J.-B., {Morris}, E., {et~al.} 2021, \aj, 162, 148,
  \dodoi{10.3847/1538-3881/ac1349}

\end{thebibliography}
\bibliographystyle{aasjournal}



\end{document}